\documentclass[a4paper, 11pt]{article} %le openany permet de commencer les chapitres sur une page paire ou impaire
\usepackage[english]{babel}
\usepackage{amsfonts}
\usepackage{amsmath}
\usepackage{amssymb}
\usepackage{array}
\usepackage{dcolumn}
\usepackage[mathcal]{eucal}
\usepackage[utf8]{inputenc}
\usepackage{geometry}
\usepackage{multicol}
\usepackage{psfrag}
\usepackage[dvips]{graphicx}
\usepackage{vmargin} %pas avec les changements directs de marges !
\usepackage[ec]{aeguill}
\usepackage{enumerate} 
\usepackage{version} %pour pouvoir faire des \begin{comment} qui contiendront plusieurs lignes
\usepackage[T1]{fontenc}
\usepackage{url}
\usepackage{indentfirst}
\usepackage{multirow}
\usepackage{setspace}
\usepackage{fancybox}
\usepackage{cases}
\usepackage{xcolor} % pour barbouiller
\usepackage[pdfborder={0 0 0}]{hyperref}
\usepackage{arydshln}
\usepackage{sectsty}
\usepackage{eurosym}

\usepackage{enumitem}  \setlist{nosep} %pour ne pas avoir d'espace autour des listes tq itemize ou enumerate

\usepackage{tikz}
\usepackage{ulem}

   \newenvironment{changemargin}[2]{\begin{list}{}{%
   \setlength{\topsep}{0pt}%
   \setlength{\leftmargin}{0pt}%
   \setlength{\rightmargin}{0pt}%
   \setlength{\listparindent}{\parindent}%
   \setlength{\itemindent}{\parindent}%
   \setlength{\parsep}{0pt plus 1pt}%
   \addtolength{\leftmargin}{#1}%
   \addtolength{\rightmargin}{#2}%
   }\item }{\end{list}}
\setmarginsrb{3cm}{2cm}{3cm}{2cm}{0cm}{1cm}{0cm}{1cm} %g, h, d, b
\newcommand{\bs}[1]{\boldsymbol{#1}}

\title{Quantum-like models\\ cannot account for the conjunction fallacy}

\author{Thomas Boyer-Kassem\footnote{Tilburg Center for Logic, General Ethics, and Philosophy of Science, Tilburg University, PO Box 90153, 5000 LE Tilburg, The Netherlands. E-mail: \href{mailto:t.c.e.boyer-kassem@uvt.nl}{t.c.e.boyer-kassem@uvt.nl}}, 
Sébastien Duchêne\footnote{GREDEG (UMR 7321: CNRS, Université de Nice Sophia Antipolis), 250 rue Albert Einstein, 06560 Valbonne, France. E-mails: \href{mailto:sebastien.duchene@gredeg.cnrs.fr}{sebastien.duchene@gredeg.cnrs.fr}, \href{mailto:eric.guerci@gredeg.cnrs.fr}{eric.guerci@gredeg.cnrs.fr}}, 
Eric Guerci\textsuperscript{$\dagger$}}

\date{\today}

\begin{document}

\maketitle

\begin{abstract}
Human agents happen to judge that a conjunction of two terms is more probable than one of the terms, in contradiction with the rules of classical probabilities---this is the conjunction fallacy. One of the most discussed accounts of this fallacy is currently the quantum-like explanation, which relies on models exploiting the mathematics of quantum mechanics. The aim of this paper is to investigate the empirical adequacy of major quantum-like models which represent beliefs with quantum states. We first argue that they can be tested in three different ways, in a question order effect configuration which is different from the traditional conjunction fallacy experiment. We then carry out our proposed experiment, with varied methodologies from experimental economics. The experimental results we get are at odds with the predictions of the quantum-like models. This strongly suggests that this quantum-like account of the conjunction fallacy fails. Future possible research paths are discussed. 
\end{abstract}

\section{Introduction}
\label{sec_intro}

Conjunction fallacy was first empirically documented by Tversky and Kahneman (1982, 1983) 
through a now renowned experiment in which subjects are presented with a description of someone called ``Linda":
\begin{quote}
``Linda is 31 years old, single, outspoken, and very bright. She majored in philosophy. As a student, she was deeply concerned with issues of discrimination and social justice, and also participated in anti-nuclear demonstrations."
\end{quote}
Then, subjects are shown a list of 8 possible outcomes describing her present employment and activities, and are asked to rank the propositions by representativeness or probability. 
Two items were specifically tested:
\begin{itemize}
\item[(1)] ``Linda is a bank teller", %proposed in the sixth position among the eight proposals,
\item[(2)] ``Linda is a bank teller and is active in the feminist movement". %, proposed in the eighth position.
\end{itemize}

Empirical results show that most people judge (2) more probable than (1). In the framework of classical probabilities, this is a fallacy---the conjunction fallacy---, since a conjunction cannot be more probable than one of its components. If Linda being active in the feminist movement is denoted by $F$ and Linda being a bank teller by $B$, then $p(F \cap B) \leqslant p(B)$ should classically prevail.

The conjunction fallacy has been shown to be particularly robust under various variations of the initial experimental protocol (cf. Tversky and Kahneman 1982, 1983, Gigerenzer 1996, Kahneman and Tversky 1996, Hertwig 1997, Hertwig and Chase 1998, Hertwig and Gigerenzer 1999, Mellers et al. 2001, Stolartz-fantino et al. 2003, Bonini et al. 2004, Tentori et al. 2004, Hertwig et al. 2008, Moro 2009, Kahneman 2011, Erceg and Galic 2014; for a review, cf. Moro 2009). %c'est Tentori et al 2013 p. 236 qui le disent
%\footnote{(Tentori Crupi Rosso 2013 donnent une liste p. 236 qui est assez différente... reprendre (aussi) celle-là ?)}
It has been observed in other cases than the Linda story, about topics like sports, politics, or natural events, and in scenarios in which the propositions to be ranked are not preceded with a description. The fallacy also persists when the experimental setting is changed, e.g. in ``between subjects" experiments in which (1) and (2) are presented to different subjects only. % (Kahneman and Tversky 1982).
Semantic and syntactic aspects have also been discussed, in relation with possible misunderstandings, like the implicit meaning of the words ``probability" and ``and". %,  that could be interpreted as a ``union" operator 
Careful experiments show that the conjunction fallacy persists.

The conjunction fallacy questions %a basic tenet of human rationality. It puts into question 
the fact that classical probability theory can be used to describe human judgment and decision making, and it can also be viewed as a challenge to the definition of what a rational judgment is.
Thus, it is no surprise that the conjunction fallacy has been the subject of a big amount of research (Tentori and Crupi 2012 give the number of a hundred papers devoted to it). It has interested psychologists, economists and philosophers alike. 
For instance, behavioral economists have looked at the consequences of the fallacy for understanding real life economic behavior, measuring the robustness of this bias in an economic context with incentives or in betting situations (e.g. Charness et al. 2010, Nilsson and Anderson 2010, Erceg and Galic 2014). They have also investigated whether the cognitive abilities of subjects are related to behavioral biases in general (and to the conjunction fallacy in particular, cf. Oechssler et al. 2009), and this has led to stimulating research with applications in finance.
Epistemologists have made confirmation and Bayesianism enter the debate (e.g. Tentori and Crupi 2008 and 2012, %), and some have questioned the ``fallacy" qualifier (
Hartmann and Meijs 2012, %Peijnenburg 2012, 
Schupbach 2012, Shogenji 2012). %vient d'un même special issue de Synthese

\bigskip

Given that a conjunction fallacy occurs under robust experimental conditions, a natural question arises: how can this fallacy be explained? 
%That is, how do human agents reason so that they are prone to a conjunction fallacy
%---or is it not a fallacy after all?
Several accounts have been argued for, but no one has reached an uncontroversial status today
(as noted by Fisk 2004, Nilsson et al 2009, Jarvstad and Hahn 2011, Tentori et al. 2013).
First, Tversky and Kahneman originally suggested that a representativeness heuristic (i.e. the probability that Linda is a feminist is evaluated from the degree with which the instance of Linda corresponds to the general category of feminists) could account for some conjunction fallacy cases. But it has been argued that the representativeness concept involved is informal and ill-specified (Gigerenzer 1996, Birnbaum et al 1990), %, even if Kahneman and Frederick 2002; for investigations around a likelihood value, see.
and suggestions to specify it in the technical sense of a likelihood value (Shafir et al 1990, Massaro 1994) account for limited cases only (Crupi et al. 2008).
According to another suggestion, agents actually evaluate the probability of the conjunction from some combination of the probabilities of the components, like averaging or adding (Fantino et al. 1997, Nilsson et al. 2009). However, such explanations do not resist empirical tests, as Tentori et al. (2013) have argued. The latter propose an account of the conjunction fallacy based on the notion of inductive confirmation as defined in Bayesian theory, and give experimental grounds for it---it is one of the currently promising accounts. 
Others have argued, also within a Bayesian framework, that there are cases in which the conjunction fallacy is actually not a fallacy and can be accounted for rationally (Hintikka 2004, von Sydow 2011, Hartmann and Meijs 2012). % Von Sydow proposed a Bayesian logic model 
Finally, another prominent proposal to account for the conjunction fallacy, on which we focus here, makes uses of so-called ``quantum-like" models, which rely on the mathematics of a major contemporary physical theory, quantum mechanics (Franco 2009, Busemeyer et al. 2011, Yukalov and Sornette 2011, Pothos and Busemeyer 2013)---note that only mathematical tools of quantum mechanics are exploited, and that the models are not justified by an application of quantum physics to the brain.

The quantum-like account of the conjunction fallacy is particularly promising as it belongs to a more general theoretical framework of quantum-like modeling in cognition and decision making, which has been applied to many fallacies or human behavior considered as irrational (for reviews, see Pothos and Busemeyer 2013, Ashtiani and Azgomi 2015, or Bruza et al. 2015; textbooks include Busemeyer and Bruza 2012, Haven and Khrennikov 2013). 
%The field of QL models is vast and is at the crossroads of economics, psychology, physics, philosophy and mathematics.
For instance, quantum-like models of judgments have been proposed to account for order effect, i. e. when the answers given to two questions depend on the order of presentation of these questions 
(Atmanspacher and Römer 2012, Busemeyer and Bruza 2012, Wang and Busemeyer 2013, Wang et al. 2014);
%Conte et al. (2009) 
for the violation of the sure thing principle, which states that if an agent prefers choosing action A to B under a specific state of the world and also prefers choosing A to B in the complementary state, then she  should choose A over B regardless of the state of the world
%. The paradoxes of Allais (1953) or Ellsberg (1961) and the disjunction effect highlighted by Shafir and Tversky (1992) show the violation of this axiom. To account for this ``irrational behavior" for classical theories, some QL models have been proposed. 
(Busemeyer et al. 2006a, Busemeyer et al. 2006b, Busemeyer and Wang 2007, Khrennikov and Haven 2009; for Ellsberg's paradox more specifically, cf. Aerts et al. 2011, Aerts and Sozzo 2013, Aerts et al. 2014; for Allais' paradox, cf. Khrennikov and Haven 2009, Yukalov and Sornette 2010, Aerts et al. 2011); %continue this work, using 
%then use the wave function to understand Ellsberg paradox. Pothos and Busemeyer (2009), Busemeyer et al.(2009) or Trueblood and Busemeyer (2011) proposed a dynamic QLmodel --- in the Hilbert space, with the Schrödinger's equation --- that rely on a time evolution of the mental state to account for some experiments described by Kahneman and Tversky. In another approach, Aerts et al. (2011) introduce the notion of ``quantum conceptual thought" to offer a QL model that illustrates the Hawaï problem.
for asymmetry judgments in similarity, i.e. that ``A is like B" is not equivalent to ``B is like A" (Pothos and Busemeyer 2011); for paradoxical strategies in game theory such as in the prisoner's dilemma (Piotrowski and Sladowski 2003, Landsburg 2004, Pothos and Busemeyer 2009, Brandenburger 2010).
%\footnote{Citer plus des papiers de Aerts, Khrenn. On est trop Busemeyer-oriented. Regarder par ex dans la revue de Ashtiani}
More generally, new theoretical frameworks with quantum-like models have been offered in decision theory and bounded rationality (Danilov and Lambert-Mogiliansky 2008 and 2010, Lambert-Mogiliansky et al. 2009, Yukalov and Sornette 2011).

As the quantum-like account of the conjunction fallacy is one of the few promising accounts of the conjunction fallacy that are discussed today, we choose to focus on it in this paper. More specifically, we focus on the class of quantum-like models which 
are presented or defended in Franco (2009), Busemeyer et al. (2011), Busemeyer and Bruza
(2012), Pothos and Busemeyer (2013) and Busemeyer et al. (2015).\footnote{There exist other quantum-like models or theories that claim to account for the conjunction fallacy, like Yukalov and Sornette (2010, 2011).} 
In these models, an agent's belief is represented by a quantum state --- and not for instance by a measurement context.
Our aim is to assess the empirical adequacy of these quantum-like models that are used to account for the conjunction fallacy. We think that two points deserve particular scrutiny. 
First, it is not always clear which version of the models are supposed to account for particular cases of conjunction fallacies---are the simplest ones, called non-degenerate, sufficient? or are the more general ones, called degenerate, needed? More recent works tend to favor degenerate models over non-degenerate ones,  and non-degenerate models have received some recent criticisms (cf. Tentori and Crupi 2013 and Pothos and Busemeyer 2013, p. 315-316), but a clear and definitive argument on the matter would be welcome.
Second, the models have not yet been much tested on other predictions than the ones they were intended to account for. It should be checked that they are not \textit{ad hoc} by testing their empirical adequacy in general.
It is understandable that these two points have not been tested beforehand, as a new general pattern of explanation for the conjunction fallacy is hard to come up with. But since the models have come to be seen as one of the most promising accounts, it becomes urgent to assess them empirically more thoroughly---this is our goal in this paper. 

As for the first point---discriminate between non-degenerate and degenerate models---, we follow a suggestion made by Boyer-Kassem et al. (2016) to test so-called ``GR equations", that are empirical predictions made by non-degenerate models\footnote{In Boyer-Kassem et al. (2016), the test is made for quantum-like order effect models.}. Such a GR test requires a new kind of experiment: not the original Linda experiment, in which agents have to rank propositions, but an order effect experiment, in which two yes-no questions are asked in one order or in the other, to different agents. 
Existing data cannot answer the question of whether the GR equations are verified, as was already noted in 2009 by Franco:
%--- who was one of the first authors to propose a QL model for conjunction fallacy --- highlighted the lack of experimental data on this question: 
\begin{quote}
``There are no experimental data on order effects in conjunction fallacy experiments, when the judgments are performed in different orders. Such an experiment could be helpful to better understand the possible judgment strategies." (Franco 2009, 421)
\end{quote}
We fill this gap here by running several order effect experiments that collect the needed data. %---the ones that Franco is talking about are exactly those needed to test the GR equations. 

As for the second point---test new empirical predictions of the models---, we consider two tests that apply to any version of the quantum-like models, whether degenerate or not, that are used in the account of the conjunction fallacy. 
It is well-known in the literature that quantum-like models that account for the conjunction fallacy predict an order effect for the two questions associated with the conjunction (``Is Linda a bank teller?" and ``Is Linda a feminist?").
Actually, this predicted order effect is not a side effect of the quantum-like models, but a core feature of them: they cannot account for the conjunction fallacy without it. This enables a direct test of the quantum-like account of the conjunction fallacy, that we apply to our collected experimental data.
Also, it has been shown that any quantum-like model of the kind involved in the account of the conjunction fallacy must make an empirical prediction called the ``QQ equality" (Wang and Busemeyer 2013, Wang et al. 2014). 
We thus test whether the QQ equality is verified. The failure of any of these last two tests will be enough to refute the current quantum-like account of the conjunction fallacy. 
Here also, the needed data is not available in the literature, but can be conveniently obtained from the same above-mentioned new experimental configuration, with two yes-no questions in both order.
Note that our methodology is novel: we are not testing the quantum-like models against data produced by traditional conjunction fallacy experiments that the model were designed to explain, but we are testing them against other data, in a new experimental framework on which the models actually make some predictions, and it is why the experimental situation we shall consider is different from the usual Linda experiment. 
Our experiment instantiates the mechanism that the quantum-like account claims agents follow: to evaluate a conjunction like ``feminist and bank teller", agents are supposed to evaluate one characteristic after another, answering for themselves to two yes-no questions (``is Linda a feminist?", ``is Linda a bank teller?"). In other words, the experiment we run somehow forces agents to follow the purported quantum-like mechanism.

To have more powerful tests, we have conducted several experiments, with variations of the scenario (Linda, but also others known as Bill, Mr. F. and K.), of the protocol (questionnaires or computer-assisted experiment) and with or without monetary incentives.
The results we obtain show that current quantum-like models are not able to account for the conjunction fallacy. %We thus suggest some possible future research paths.

The outline of the paper is the following. In Section 2, a general quantum-like model is introduced.
Section~3 presents the three empirical tests that will be performed: the GR equations, order effect, and the QQ equality. The experimental protocol is presented in Section~4, and the results in Section~5. Section~6 presents the statistical analysis, and Section~7 discusses the scope of the results and the future of the research on the conjunction fallacy account.

\section{A quantum-like account of the conjunction fallacy}
\label{sec_QL-models}

As indicated in the introduction, we focus in this paper on a family of quantum-like models based on similar hypotheses that have recently been proposed to account for the conjunction fallacy. %and to account for the empirical results of Linda's experiment, 
They are presented or defended in Franco (2009), Busemeyer et al. (2011), Busemeyer and Bruza (2012), Pothos and Busemeyer (2013) and Busemeyer et al. (2015).%
\footnote{There exist other quantum-like models or theories that claim to account for the conjunction fallacy, like Yukalov and Sornette (2010, 2011). However, the latter theory does not display some features that are central to our present tests (like the reciprocity law), which casts doubt on the possibility to test it in the same way.}
For simplicity, we choose here to summarize them with a single model with our own notations, and the correspondence with the various models from the literature can easily be made by the reader.
For illustrative purposes, we shall consider the conjunction fallacy through the Linda case, but the generalization to other instances of the conjunction fallacy are straightforward.

According to this literature, after reading Linda's description, the subject who has to choose the more likely proposition between 
\begin{itemize}
\item[(1)] ``Linda is a bank teller",
\item[(2)] ``Linda is a feminist and a bank teller".\footnote{The original sentence used in Tversky and Kahneman (1983) is now abridged in this form, as robustness studies have shown that the existence of the fallacy does not depend on such details.}
\end{itemize}
has the following mental process.
To compare the propositions, she evaluates each one in terms of a yes-no question:
\begin{itemize}
\item[($Q_1$)] ``Is Linda a bank teller?",
\item[($Q_2$)] ``Is Linda a feminist and a bank teller?".
\end{itemize}
An important hypothesis of the quantum-like models is that, when the subject considers ($Q_2$), she actually answers for herself successively two simple yes-no questions:
\begin{itemize}
\item[($Q_F$)] ``Is Linda a feminist?",
\item[($Q_B$)] ``Is Linda a bank teller?".
\end{itemize}
Answering ``yes" to $Q_2$ amounts to answering ``yes" to both $Q_F$ and $Q_B$.
Also, the hypothesis is made that the more probable outcome (bank teller or feminist) is evaluated first. As the description of Linda makes her more likely a feminist than a bank teller,  this means that $Q_2$ is answered by answering first $Q_F$ and then $Q_B$.\footnote{Franco (2009) does not explicitly make this hypothesis, but he implicitly considers that the conjunction (2) will be evaluated by answering $Q_F$ \textit{and then} $Q_B$ (p. 418). 
Anyway, the tests we consider in the forthcoming sections do not depend on this hypothesis.}
Let us now turn to the quantum-like framework that enable the quantitative prediction of the conjunction fallacy, $p(2)>p(1)$.

\subsection{Quantum-like models}
\label{subsec_QL-models}

For pedagogical purposes, the non-degenerate versions of the quantum-like models are presented first, and the degenerate versions afterwards.
The belief states of agents are represented within a vector space. In the simple case where an agent have just given an answer ``yes" (respectively ``no") to question $Q_F$, her belief state is represented by the vector $\bs{F_y}$ (respectively, $\bs{F_n}$). In accordance with the literature, we shall say for short that these vectors represent the answers themselves.
Similarly with $\bs{B_y}$ and $\bs{B_n}$ for answers to question $Q_B$. 
The sets ($\bs{B_y}$, $\bs{B_n}$) and ($\bs{F_y}$, $\bs{F_n}$) respectively represent all possible answers to questions $Q_B$ and $Q_F$, and thus each one is a basis of the same 2-dimension vector space. 

\begin{figure}[!ht]
\begin{center}
\setlength{\unitlength}{.7mm}
% %
% schéma de gauche
\begin{picture}(100,80)(-30,-10) %(taille)(coord du coin inférieur gauche)
%a_0
\put(0, 0){\vector(1,0){50}}
\put(52, -2){$\bs{B_y}$}
%a_1
\put(0, 0){\vector(0,1){50}}
\put(-5, 55){$\bs{B_n}$}
%b_0
\put(0, 0){\vector(3,1){47.4}} % 50*cos(arctan(y/x))
\put(50, 15){$\bs{F_y}$}
%b_1
\put(0, 0){\vector(-1,3){15.8}} % 50*cos(arctan(y/x))
\put(-25, 50){$\bs{F_n}$}
\end{picture}
\qquad
\begin{picture}(100,80)(-30,-10) %(taille)(coord du coin inférieur gauche)
%    \put(0,0){\line(1,0){120}}
%    \thicklines 
%a_0
\put(0, 0){\vector(1,0){50}}
\put(52, -2){$\bs{B_y}$}
%a_1
\put(0, 0){\vector(0,1){50}}
\put(-5, 55){$\bs{B_n}$}
%b_0
\put(0, 0){\vector(3,1){47.4}} % 50*cos(arctan(y/x))
\put(50, 15){$\bs{F_y}$}
%b_1
\put(0, 0){\vector(-1,3){15.8}} % 50*cos(arctan(y/x))
\put(-25, 50){$\bs{F_n}$}
%angle
%\qbezier(30,0)(30,5)(28.4,9.4)
%\put(32,3){$\beta$}
\thicklines 
%psi
\put(0, 0){\vector(4,3){40}} % 50*cos(arctan(y/x)) %note : c'est pratique, l'angle entre a0 et b0 est égal à celui entre b0 et psi
\put(42, 33){$\bs{\Psi}$}
\thinlines
%pointillés : proj de psi sur a0 et a1
\qbezier[30](40,0)(40,15)(40,30)
\put(30, -7){$\bs{B_y} \cdot \bs{\Psi}$}
\qbezier[40](0,30)(20,30)(40,30)
\put(2, 33){$\bs{B_n} \cdot \bs{\Psi}$}
%pointillés : proj de psi sur b0 et b1
\qbezier[10](45,15)(42.5,22.5)(40,30)
%\put(44, 9){$\beta_0$}
\qbezier[30](-5,15)(17.5,22.5)(40,30)
%\put(-12, 10){$\beta_1$}
\end{picture}
\end{center}
\caption{[Left:] The two bases corresponding to the answers ``yes" and ``no" to questions $Q_B$ and $Q_F$.
[Right:] The state vector $\bs{\Psi}$ can be decomposed on the two orthonormal bases (the scalar products on $\bs{B_y}$ and $\bs{B_n}$ are indicated). These figures assume the special case of a Hilbert space on real numbers.}\label{fig_bases-B-and-F}
\end{figure}
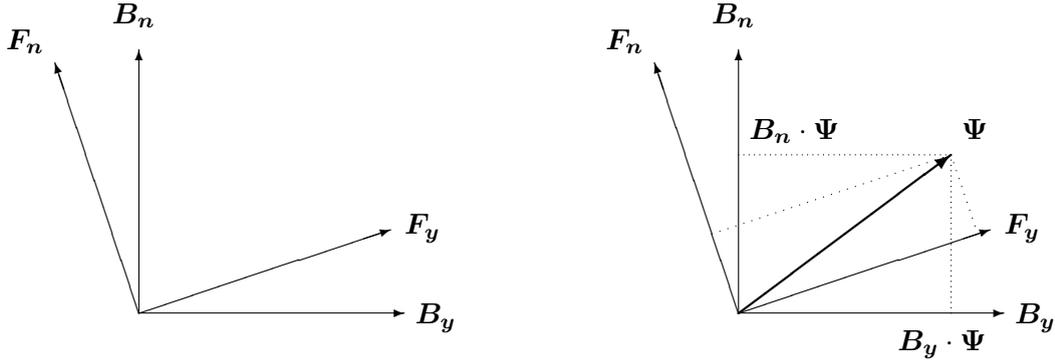

The vector space is equipped with a scalar product, thus becoming a Hilbert space: for two vectors $\bs{W}$ and $\bs{X}$, the scalar product $\bs{W} \cdot \bs{X}$ is a complex number. The order of the vectors within a scalar product here matters: $\bs{X} \cdot \bs{W}$ is the complex conjugate of $\bs{W} \cdot \bs{X}$.
The above bases are supposed to be orthogonal: $\bs{B_y} \cdot \bs{B_n} =  \bs{F_y} \cdot \bs{F_n} = 0$, and of unitary norm:  $\bs{B_y} \cdot \bs{B_y} =  \bs{B_n} \cdot \bs{B_n} = \bs{F_y} \cdot \bs{F_y} =  \bs{F_n} \cdot \bs{F_n} = 1$.
A representation of the bases in the special case of real coefficients can be found on Figure~\ref{fig_bases-B-and-F} [Left].

An agent's state of belief is represented by a normalized vector $\bs{\Psi}$ within the Hilbert space. This vector can be decomposed in either of the two above-mentioned bases, as indicated on Figure~\ref{fig_bases-B-and-F} [Right]: 
\begin{equation}
\bs{\Psi} = (\bs{B_y} \cdot \bs{\Psi}) \bs{B_y} 
+(\bs{B_n} \cdot \bs{\Psi}) \bs{B_n} 
= (\bs{F_y} \cdot \bs{\Psi}) \bs{F_y} 
+(\bs{F_n} \cdot \bs{\Psi}) \bs{F_n}.
\end{equation}
With the specific values taken in Figure~\ref{fig_bases-B-and-F} [Right] in a Hilbert space on real numbers, this equation becomes for instance:
\begin{equation}
\bs{\Psi} = 0.8 \bs{B_y} 
+0.6 \bs{B_n} 
\approx 0.949 \bs{F_y} 
+ 0.316 \bs{F_n}.
\end{equation}

The belief state $\bs{\Psi}$ gathers all the relevant information needed to predict the behavior of the agent, in the following way. Predictions made by the quantum-like models are probabilistic. When a question $Q_X$ ($X$ = $B$ or $F$) is asked, the probability that the agent answers $X_i$ (\mbox{$i$ = $y$ or $n$}) is given by the squared modulus of the scalar product between the belief state and the vector representing the answer:
\begin{equation}\label{eq_Born-rule}
p(X_i) =  \vert \bs{X_i} \cdot \bs{\Psi} \vert^2.
\end{equation}
This rule is usually called the Born rule, in analogy with the quantum mechanics denomination.
It enables to compute the probability that the agent gives each of the 4 answers, in case questions $Q_B$ or $Q_F$ are asked (as $\bs{\Psi}$ is normalized, $p(X_y) + p(X_n) = 1$).
In the case of a real Hilbert space like on Figure~\ref{fig_bases-B-and-F}, a geometric interpretation of the Born rule is the following: to compute the probability to answer, say, ``yes" to question $Q_B$, orthogonally project $\bs{\Psi}$ on $\bs{B_y}$ --- this gives the length $\bs{B_y} \cdot \bs{\Psi}$, and the wanted probability is just the squared of it. So, the more $\bs{\Psi}$ is aligned with a basis vector $\bs{X_i}$, the larger the probability is that the agent will answer $i$ if question $Q_X$ is posed (note the ``if question $Q_X$ is posed" part: in quantum-like models, the probability of an answer is only defined in the context in which the corresponding question is posed). % Thus, when question B has just been posed, you can say that the probability that the agent will answer ``yes" is 80 \%, but you can't say that in this context the probability that the agent will answer ``yes" to $Q_F$ is of 90 \%.
For instance, with the specific values in Figure~\ref{fig_bases-B-and-F} [Right], $p(B_y) = 0.64$, $p(B_n) = 0.36$, $p(F_y) = 0.9$ and $p(F_n) = 0.1$, which is consistent with the relative alignments of the basis vectors with $\bs{\Psi}$.
%As the state vector is more aligned with $\bs{B_y}$ than $\bs{B_n}$ for instance, $p(B_y) > p(B_n)$.}

The last postulate of the quantum-like model has to do with the way $\bs{\Psi}$ changes over time. First, $\bs{\Psi}$ does not change unless the agent answers a question. This conveys the fact that the agent's beliefs are not externally influenced. This hypothesis is supposed to be relevant for cases in which the questions are posed to the agent relatively quickly. 
Second, when the agent answers a question $Q_B$ or $Q_F$, her state of belief changes. If her answer to question $Q_X$ is $X_i$, then her new state of belief just after giving the answer is: 
\begin{equation}\label{eq_projection-postulate}
\bs{\Psi} \quad \longmapsto 
\quad \frac{\bs{X_i} \cdot \bs{\Psi}}{\vert \bs{X_i} \cdot \bs{\Psi} \vert}  \bs{X_i}.
\end{equation}
As the fraction in Eq.~\ref{eq_projection-postulate} is a complex number, the state of belief after an answer $X_i$ is proportional to the vector $\bs{X_i}$ representing this answer. In the case of a real Hilbert space like on Figure~\ref{fig_bases-B-and-F}, after answering ``yes" to question $Q_B$, $\bs{\Psi}$ becomes either $\bs{B_y}$ or $-\bs{B_y}$, %(depending on the relative orientation of the vectors before the q
whatever the state of belief before the question. In other words, after a question $X$ has been posed, the state of belief is bound to be along the basis vectors representing its answers.
Eq.~\ref{eq_projection-postulate} can be interpreted as follows: the $(\bs{X_i} \cdot \bs{\Psi}) \bs{X_i}$ part represents the fact that $\bs{\Psi}$ is projected on $\bs{X_i}$, the basis vector representing the given answer; the $1/ \vert \bs{X_i} \cdot \bs{\Psi} \vert$ part is then just a multiplicative factor that ensures that the new state of belief is normalized. Hence, the above rule is often called the projection postulate.

%As the state of belief is changes as an answer is given, 
Because of the projection postulate, the states before and after an answer are in general different. They are the same only if the state previous to the answer is proportional to one of the basis vectors representing the possible answers to the question, i. e. when $\bs{\Psi} = \lambda  \bs{X_i}$, where $\lambda$ is a complex number such that $\vert \lambda \vert = 1$ (in the real case, $\bs{\Psi} = \pm \bs{X_i}$). In such a case, the agent answers $i$ to question $X$ with probability 1, and Eq.~\ref{eq_projection-postulate} states that $\bs{\Psi} \quad \longmapsto \quad \bs{X_i}.$
The fact that the state of belief changes when a question is answered is a real departure from the classical viewpoint. Classically, the answer is supposed to \textit{reveal} a belief, which is pre-existent to the question, and is the same before and after. However, the quantum-like models predict that once a question has been answered, the same answer will be given if the same question is posed again just after. 

%One can note that, in this theoretical framework, answering a first question modifies the state of the agent, and thus order effect can occur (e.g when  p($F_y$ then $B_y$) $\neq$ p($B_y$ then $F_y$).\\

\bigskip
Let us now turn to the more general versions of these models, the degenerate ones. The difference lies in the fact that an answer is not represented by a vector belonging to a 1D space, but by any subspace of dimension $m$, for instance a plane. Then, the Hilbert space is not of dimension 2, but of a higher one.
When question $Q_X$ is posed, the probability that the agent answers $X_i$ is now defined as:
\begin{equation}
p(X_i) =  \vert P_{X_i} \cdot \bs{\Psi} \vert^2.
\end{equation}
where $P_{X_i}$ is the orthogonal projector onto the subspace representing answer $i$ to question $Q_X$. The change in the state of belief is now:
\begin{equation}
\bs{\Psi} \quad \longmapsto 
\quad \frac{P_{X_i} \cdot \bs{\Psi}}{\vert P_{X_i} \cdot \bs{\Psi} \vert}.
\end{equation}
For the rest, the model is the same.

\subsection{Accounting for the fallacy}
\label{subsec_AccFallacy}

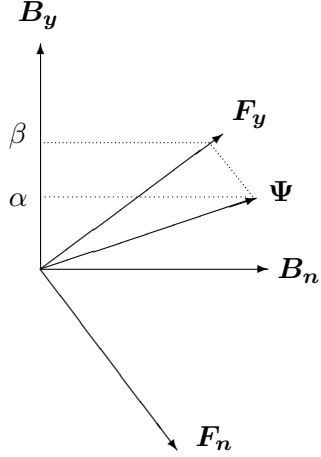
\begin{figure}[!tb]
\begin{center}
\setlength{\unitlength}{.6mm}
\begin{picture}(90,90)(-10,-30) %(taille)(coord du coin inférieur gauche)
%    \put(0,0){\line(1,0){120}}
%    \thicklines 
%B_n
\put(0, 0){\vector(1,0){50}}
\put(52, -2){$\bs{B_n}$}
%B_y
\put(0, 0){\vector(0,1){50}}
\put(-5, 55){$\bs{B_y}$}
%psi
\put(0, 0){\vector(3,1){47.4}} % 50*cos(arctan(y/x))
\put(50, 15){$\bs{\Psi}$}
%F_n
\put(0, 0){\vector(3,-4){30}} % 50*cos(arctan(y/x))
\put(34, -39){$\bs{F_n}$}
%\thicklines 
%F_y
\put(0, 0){\vector(4,3){40}} % 50*cos(arctan(y/x)) %note : c'est pratique, l'angle entre a0 et b0 est égal à celui entre b0 et psi
\put(42, 33){$\bs{F_y}$}
\thinlines

\put(-7, 28){$\beta$}
\qbezier[20](47,15.5)(42,22)(37,28)
\qbezier[45](37,28)(18.5,28)(0,28)
\qbezier[45](45,16)(22.5,16)(0,16)

\put(-7, 14){$\alpha$}

\end{picture}
\end{center}
\caption{A quantum-like account of the conjunction fallacy in Linda's scenario. This figure assumes the special case of a Hilbert space on real numbers.}\label{fig_bases-QP-explanation}
\end{figure}

The mental process that gives rise to the conjunction fallacy that has been described at the beginning of this Section is graphically illustrated
%First, when an agent considers the conjunction ``Linda is a bank teller and a feminist", this conjunction is evaluated as a sequence of projections, corresponding to the answers to two successive and dichotomous yes-no questions, namely taken among the set $\{F, B\}$. Secondly, the actual description of Linda ``makes it very likely that she is feminist, very unlikely that she is a bank teller". Thirdly, the hypothesis is made that ``the more probable possible outcome is evaluated first" --- in other words, question $Q_F$ is evaluated before question $Q_B$. 
%Thus the authors suggest a quantum model with order effects --- in the same way as quantum models for order effects --- to account for conjunction fallacy.
on Figure~\ref{fig_bases-QP-explanation}. The probability of considering that Linda is a bank teller corresponds to the squared length of the projection of $\bs{\Psi}$ onto the bank teller vector $\bs{B_y}$, and $p(B) = |\alpha|^2$. For instance, with the specific values used in Figure~\ref{fig_bases-QP-explanation} with a real Hilbert space, $\alpha \approx 0.316$ and $p(B) = 0.1$.
On the other hand, the probability of considering her to be feminist and bank teller corresponds to the squared length of the projection of $\bs{\Psi}$ onto two successive vectors, first $\bs{F_y}$ and then $\bs{B_y}$, and $p(F \cap B) = |\beta|^2$. In the example of Figure~\ref{fig_bases-QP-explanation}, $\beta = 0.6$ and $p(F \cap B) = 0.36$.

So, there exist some model configurations, like the one plotted on Figure~\ref{fig_bases-QP-explanation}, in which the probability to be judged feminist and bank teller is higher than the probability to be judged bank teller, leading to
\begin{equation}
p(F \cap B) > p(B),
\end{equation}
in accordance with empirical results. 
A quantum-like model of the conjunction fallacy has been provided.%
\footnote{We have slightly simplified the account given by Busemeyer et al. (2011). When an agent evaluates the conjunction, they do suppose that answering for herself the first question $Q_F$ projects her state vector onto either $\bs{F_y}$ or $\bs{F_n}$, but they do not suppose that answering for herself the second question $Q_B$ projects the state vector onto $\bs{B_y}$ or $\bs{B_n}$, because they argue that what is needed at this time is only an evaluation of the probability, and not a firm answer (the authors acknowledge the validity of the projection postulate as soon as the agent gives a definite answer to a question). So, the authors actually do not specify what the state vector is after the evaluation of the conjunction (personal communication, 2014). For simplicity, we have made as if the state vector was projected onto either $\bs{B_y}$ or $\bs{B_n}$, like for other questions, but this has no consequence for our forthcoming tests. See also Section~\ref{sec_expdesign}.}

\section{Empirical tests}
\label{sec_empirical-tests}

This section presents the three empirical predictions of the above quantum-like model that we will test. The first one applies to non-degenerate models, while the others apply to non-degenerate and degenerate models.

\subsection{The GR equations}\label{sec:GReq}

%From the quantum-like model presented in Section~\ref{subsec_QL-models}, 
Following Boyer-Kassem et al. (2016), some specific empirical predictions can be derived for non-degenerate models, i.e. in which the answers are represented by subspaces of dimension~1. %, that have established these predictions in the case of a more general quantum-like model. 
It can be shown that a well-known law from quantum mechanics, the law of reciprocity, holds. 
Consider the two questions $Q_F$ and $Q_B$ in one order or in the other. The law of reciprocity states that, for $(X, Y) \in \{B,F\}^2$, and $(i, j) \in \{y, n\}^2$,
\begin{equation}
p(Y_j|X_i) = p(X_i|Y_j).
\end{equation}
This law asserts that conditional probabilities of an answer given another answer are the same whatever the order of the questions $Q_B$ and $Q_F$. 
Note that this law is typically quantum: it is not true in general for a classical model, in which $p(Y_j|X_i) = p(X_i|Y_j)\times p(Y_j)/ p(X_i)$, and thus $p(Y_j|X_i) \neq p(X_i|Y_j)$ as soon as $p(Y_j) \neq p(X_i)$. 
%For the model of Section~\ref{subsec_QL-models}, this 

The law of reciprocity %is quite well-known in quantum-like modeling\footnote{\label{note_refs-law-of-reciprocity}Cf. %Busemeyer (2008, 25 and 29), 
can be instantiated in the following ways:

\begin{numcases}
%{\textrm{(*)}}
\strut  
\label{eq_pb0a0=pa0b0}
%p_\textrm{\textsc{i}}(C|A) = p_\textrm{\textsc{ii}}(A|C)
p(B_y|F_y) = p(F_y|B_y),\\
\label{eq_pb1a0=pa0b1}
%p_\textrm{\textsc{i}}(D|A) = p_\textrm{\textsc{ii}}(A|D)
p(B_n|F_y) = p(F_y|B_n), \\
\label{eq_pb0a1=pa1b0}
%p_\textrm{\textsc{i}}(C|B) = p_\textrm{\textsc{ii}}(B|C)
p(B_y|F_n) = p(F_n|B_y),\\
\label{eq_pb1a1=pa1b1}
%p_\textrm{\textsc{i}}(D|B) = p_\textrm{\textsc{ii}}(B|D)
p(B_n|F_n) = p(F_n|B_n).
\end{numcases}

Some easy computation enable to show that the following equations, called the Grand Reciprocity (GR) equations, hold (cf. Boyer-Kassem et al. 2016, Section~3.1):
%\footnote{\textbf{J'ai choisi de ne pas reproduire la démonstration, qui appelle de nombreux commentaires comme dans le premier papier, et qui finit par être un longue. Cela évite d'être comme Khrennikov, qui se recopie tout le temps dans ses livres/papiers, et je trouve ça pénible. Mais on peut en discuter.}}

\begin{numcases}
%{\textrm{(***)}}
\strut  \label{eq_mix0011}
%p_\textrm{\textsc{i}}(C|A) = p_\textrm{\textsc{ii}}(A|C) = p_\textrm{\textsc{i}}(D|B) = p_\textrm{\textsc{ii}}(B|D)
%\boxed{
p(B_y|F_y) = p(F_y|B_y) = p(B_n|F_n) = p(F_n|B_n),\\
%}
\label{eq_mix1001}
%p_\textrm{\textsc{i}}(D|A) = p_\textrm{\textsc{ii}}(A|D) = p_\textrm{\textsc{i}}(C|B) = p_\textrm{\textsc{ii}}(B|C)
%\boxed{
p(B_n|F_y) = p(F_y|B_n) = p(B_y|F_n) = p(F_n|B_y).
%}
\end{numcases}
These equations \ref{eq_mix0011} and \ref{eq_mix1001} are equivalent to one another and to the law of reciprocity itself.\footnote{Boyer-Kassem (2016, Section~3) show that the GR equations are equivalent to double stochasticity in both orders (cf. also Khrennikov 2010, p. 24 and 36), and presents generalizations of the GR equations.} %, and comments on the link between these equations and other concepts like double-stochasticity, cf. Boyer-Kassem et al. (2015), Section~3. In particular, note that }
%These GR equations can be shown to hold for generalized versions of the model presented in Section~\ref{subsec_QL-models} --- for a population of $N$ agents, for agents described with mixed states, and so on. The reader is referred to Boyer-Kassem et al. (2015), Section~3.2 for details.} 
They state that the conditional probabilities that exist when $Q_B$ is asked before $Q_F$ is asked --- call it situation ($Q_B, Q_F$) --- and in the ($Q_F, Q_B$) situation are actually much constrained: among the eight quantities that can be experimentally measured, there is just one free real parameter. In other words, the non-degenerate quantum-like model presented in Section~\ref{subsec_QL-models} actually leaves very little freedom to conditional probabilities.

%To the best of our knowledge, t
The fact that the conditional probabilities are constrained by the GR equations had not been noticed beforehand for quantum-like models for the conjunction fallacy. %Our bottom line in this paper is to experimentally test these empirical claims the model makes.
Note that these empirical predictions are \textit{consequences} of the quantum-like models that are used to explain the conjunction fallacy in the Linda experiment, and that these consequences are observable in experimental situations --- $(Q_B, Q_F)$ and $(Q_F, Q_B)$ situations --- that are \textit{not} the ones of the original Linda experiment. In other words, the GR equations show that a non-degenerate quantum-like model that is used to explain a Linda experiment can be further tested on another kind of experiment. We shall come back on this point in Section~\ref{sec_expdesign}. %This is what the next Section considers.

The interpretation of the conditional probabilities is clear: they have been defined as the probability of some answer to a second question given the answer to a first question. This is straightforwardly consistent with the models presented in Section~\ref{sec_QL-models}, and in accordance with %the whole quantum-like literature on the matter, and 
classical order effect experiments. Another interpretation of the conditional probabilities could be that of an answer given some new piece of evidence, but this is not what is considered in this paper.

\subsection{Order effect}

\begin{figure}[!tbh]
\begin{center}
\setlength{\unitlength}{.7mm}
\begin{picture}(80,70)(-30,-10) %(taille)(coord du coin inférieur gauche)
%    \put(0,0){\line(1,0){120}}
%    \thicklines 
%a_0
\put(0, 0){\vector(1,0){50}}
\put(52, -2){$\bs{B_y}$}
%a_1
\put(0, 0){\vector(0,1){50}}
\put(-5, 55){$\bs{B_n}$}
%b_0
\put(0, 0){\vector(3,1){47.4}} % 50*cos(arctan(y/x))
\put(50, 15){$\bs{F_y}$}
%b_1
\put(0, 0){\vector(-1,3){15.8}} % 50*cos(arctan(y/x))
\put(-25, 50){$\bs{F_n}$}
%angle
%\qbezier(30,0)(30,5)(28.4,9.4)
%\put(32,3){$\beta$}
\thicklines 
%psi
\put(0, 0){\vector(4,3){40}} % 50*cos(arctan(y/x)) %note : c'est pratique, l'angle entre a0 et b0 est égal à celui entre b0 et psi
\put(42, 33){$\bs{\Psi}$}
\thinlines
%pointillés : proj de psi sur a0, puis sur b0
\qbezier[40](40,0)(40,15)(40,30)
\qbezier[16](40,0)(38,6)(36,12)
%pointillés : proj de psi sur b0, puis sur a0
\qbezier[10](45,15)(42.5,22.5)(40,30)
\qbezier[10](45,15)(45,7.5)(45,0)
%\put(44, 9){$\beta_0$}
\end{picture}
\end{center}
\caption{The state vector $\bs{\Psi}$, projected first on $\bs{B_y}$ and then on $\bs{F_y}$, or first on $\bs{F_y}$ and then on $\bs{B_y}$, gives different lengths. Consequently, the corresponding probabilities of answering ``yes" to questions $Q_B$ and $Q_F$ depend on the order of presentation of the questions: it is an order effect.}\label{fig_order-effect}
\end{figure}
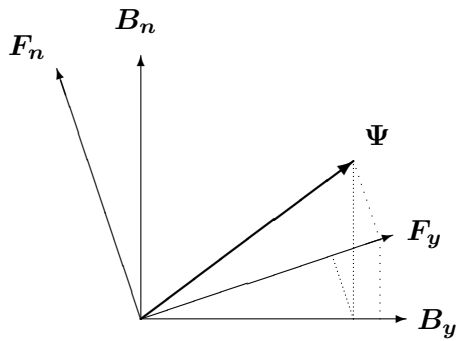

Quantum-like models of Section~\ref{subsec_QL-models} can predict an order effect, that is, predict that agents give different answers to the question $Q_F$ followed by question $Q_B$, and to the question $Q_B$ followed by question $Q_F$ (cf.~Figure~\ref{fig_order-effect}). 
This comes from the projection postulate that modifies the state of belief when an answer is given to a question. 
This order effect property of the quantum-like models is well-known, and it has actually been used to provide a quantum-like account of order effect (see for example Conte et al. 2009, Busemeyer et al. 2009, Busemeyer et al. 2011, Atmanspacher and Römer 2012, Pothos and Busemeyer 2013, Wang and Busemeyer 2013 and Wang et al. 2014, Boyer-Kassem et al. 2016) --- thus, the same models are at the basis of the account of order effect and of the conjunction fallacy.

More importantly, it can be shown that only models that display an order effect are able to account for the conjunction fallacy (cf. Busemeyer et al. 2011, Busemeyer and Bruza 2012, Bruza et al. 2015 p. 388, Busemeyer et al. 2015). In other words, the quantum-like models of Section~2 that do not present an order effect
%---for instance, a degenerate model in a 4-dimensional Hilbert space with basis vectors, $\bs{B_yF_y}$  (where $\bs{XY_{ij}}$ denotes the ), in which the answer to a question is represented by a 2D plan
cannot predict $p(F \cap B) > p(B)$, and thus cannot account for the conjunction fallacy.
%\footnote{
The reason is, in short, the following: questions $Q_B$ and $Q_F$ are either compatible or incompatible in the standard quantum sense. In the latter case, the Hilbert space is (in the simplest case) 2D, with basis vectors like on Figure~\ref{fig_bases-B-and-F}, and there is an order effect. In the former case, %which has not been considered in Section~2, 
the Hilbert space is (in the simplest case) 4D, with basis vectors ($\bs{BF_{yy}}$, $\bs{BF_{yn}}$, $\bs{BF_{ny}}$, $\bs{BF_{nn}}$), where the vector $\bs{BF_{ij}}$ stands for answer $i$ to question $Q_B$ and answer $j$ to question $Q_F$, in whatever order. And such a model displays no order effect: whatever the order of the questions, the probability of an answer $i$ to question $Q_B$ and of an answer $j$ to question $Q_F$ will be $\vert \Psi_{ij} \vert^2$, where $\Psi_{ij}$ is the coordinate along the $\bs{BF_{ij}}$ vector ($\Psi_{ij} = \bs{BF_{ij}} \cdot \Psi$).
%In this sense, the behavior is classical and does not necessitate a quantum-like model.
Can such a model predict a conjunction fallacy to occur? On the one side, consider the evaluation of the conjunction: the agent first considers $Q_F$; if she answers ``yes", the state vector is projected onto the plane $(\bs{BF_{yy}}, \bs{BF_{ny}})$. If she now answers ``yes" to $Q_B$, the resulting vector is projected onto $\bs{BF_{yy}}$. So, the probability to answer ``yes" to both questions is given by the square modulus of the $\bs{BF_{yy}}$ component, i.e. $\vert \Psi_{yy} \vert^2$. On the other side, consider the evaluation of $B$, for which the agent considers $Q_B$. If she answers ``yes", the state vector is projected onto the plane $(\bs{BF_{yy}}, \bs{BF_{yn}})$. The probability of such an answer is given by the squared modulus of the length of this projection, namely $\vert \Psi_{yy} \vert^2 + \vert \Psi_{yn} \vert^2$ (remember that the basis vectors are orthogonal). This quantity is at least larger than $\vert \Psi_{yy} \vert^2$, so a conjunction fallacy cannot occur.%}

To sum up, any quantum-like model of the kind considered in Section~2  which claims to account for the conjunction fallacy, be it non-degenerate or degenerate, has to display an order effect on the corresponding questions. This provides our second test (cf. Section~\ref{StatAnalysis} for a discussion of the mathematical expression of the test).
The proponents themselves of the quantum-like account of the conjunction fallacy consider that the use of incompatible concepts (or questions) is the key feature of their model. As incompatible questions straightforwardly imply an order effect, our order effect test is actually a direct test of the core feature of the quantum-like account.\footnote{One could consider to test whether the questions $Q_F$ and $Q_B$ are compatible or incompatible. But the easiest way to do so is actually to test the order effect on these two questions.} As for the GR equations, note that the order effect is here understood as an experimental situation with two successive yes-no questions, posed in one order or in the other after a text has been read, and that no new piece of evidence is provided between the two questions.
To sum up, three features are essential for the quantum-like models under study to account for the conjunction fallacy: the Born rule (eq.~\ref{eq_Born-rule}), the projection postulate (eq.~\ref{eq_projection-postulate}), and the presence of incompatible questions entailing order effects.

\subsection{The QQ equality}

The quantum-like models of Section~2, % that have been used to account for order effect too, 
whether degenerate or not, have recently been shown to entail new testable empirical predictions %, in addition to the order effect 
(Wang and Busemeyer 2013): a ``Quantum Question" (QQ) equality. %and Wang et al. (2014)
Noting $p(X_i, Y_j)$ the probability  of answering first $i$ to question $Q_X$ \textit{and then} $j$ to question $Q_Y$ (this is a joint probability, not a conditional probability), the QQ equality reads:
\begin{equation}
p(F_y, B_n) + p(F_n, B_y)  = p(B_y, F_n) + p(B_n, F_y).
\end{equation}

This equality is of prime importance. 
As Busemeyer et al. (2015, 241) put it, ``it is an a priori, \textit{precise}, \textit{quantitative}, and \textit{parameter-free} prediction about the pattern of order effects".
It has served as a test of the quantum-like models that claim to account for order effect. It turns out that ``it has been statistically supported across a wide range of 70 national field experiments (containing 651 to 3,006 nationally representative participants per field experiment) that examined question-order effects (Wang et al., 2014)" (\textit{ibid.}).
Similarly, the QQ equality can be empirically tested in the case of the quantum-like models that account for the conjunction fallacy, as the models are the same. This constitutes our third test (further statistical details about the test are given in Section~6).

\section{Experimental design}
\label{sec_expdesign}

The three tests presented in the previous section (GR equations, order effect, QQ equality) require to carry out an order effect experiment that shows the description of Linda and then asks the questions $Q_F$ and $Q_B$ in both orders, ($Q_F, Q_B$) or $(Q_B, Q_F)$. The former order somehow forces the agent to follow the cognitive process supposed by the quantum-like models when evaluating a conjunction. We propose here its first experimental realization, in order to test the quantum-like models of Section~2.

The order effect experiment we are considering here is different from the original conjunction fallacy experiment. If we want to claim that it tests anyway the quantum-like account of the conjunction fallacy, do we need to make some extra hypothesis? For instance, do we need to suppose that the quantum-like model for the conjunction fallacy also applies to another kind of experiment?   %that the quantum-like explanation that described can be tested in this way, or 
Or do we need to assume that forcing an agent to explicitly answer the two questions will give the same results as when she answers them for herself? We need not, because these assumptions are already made in the papers we are considering. 
First, the simple fact that the quantum-like account of the conjunction fallacy relies on ``models" that have a general and universal form\footnote{E.g. ``In general, a person's state of beliefs about the presence or absence of various feature combinations is represented by..." (Busemeyer et al 2015, p. 237).}
%(; Busemeyer and Bruza (2012, p. 108) insist on presenting a ``general quantum model" from )
and not only on \textit{ad hoc} rules that apply to a limited number of situations, allows anyone to use these models \textit{ad libitum} in any experimental situation that the model may represent. The order effect situation, in which two questions are asked, clearly falls within that range. So, we are allowed to apply (and thus to test) the quantum-like models of the conjunction fallacy in an order effect experiment. This amounts to testing experimental predictions of the models that they make because they have a general form. 
As the proponents of the models write: ``The basic quantum model underpinning the conjunction fallacy [...] makes new \textit{a priori} predictions. Foremost among them is the consequence that incompatible judgments and decisions must entail order effects." (Bruza et al. 2015, p. 388). %``The basic quantum model underpinning the conjunction fallacy also readily explains the disjunction fallacy. Furthermore, this model makes new \textit{a priori} predictions. Foremost among them is the consequence that incompatible judgments and decisions must entail order effects."
(Recall that incompatible judgments are required in the quantum-like model of the conjunction fallacy.) In other words, the conjunction fallacy model entails order effects, and thus can be tested on them.
This is all the more true than the authors actually claim that the quantum-like models used for the conjunction fallacy are the same as those used to explain other fallacies or phenomena, like order effect itself or similarity judgments. All models belong to a family that are often called a ``theory" of quantum cognition, and they are meant to make predictions on a wide range of phenomena, in diverse experimental situations --- and the authors rightly claim that this is a strength of their approach. 
This supports the generality of the quantum-like models used for the conjunction fallacy. Thus, it is legitimate to use them in other situations like the order effect one. Besides, these models \textit{have been} applied to question order effect (Wang and Busemeyer 2013, Wang et al. 2014), and it is clear that no extra hypothesis than the ones presented in Section~2 is needed for that.
In sum, the literature claims that the very same models can be used for the conjunction fallacy and for question order effect, so we are justified in testing them on new order effect cases as Linda's.

Finally, recall that we consider here two successive yes-no questions, asked in both orders. Thus, the conditional probabilities are interpreted as probabilities of a second answer given a first answer. This is fully in line with the models of the conjunction fallacy themselves. Consider for instance: ``In this problem there are two questions: the feminism question and the bank teller question. For each question, there are two answers: yes or no." (Busemeyer and Bruza 2012, p. 15); ``we consider two dichotomous questions A and B, as for example A: Is Linda a feminist? and B: Is Linda a bank teller?" (Franco 2009 p.~416 ). %``QP [quantum probability theory, i.e. the quantum-like models] defines conjunction between incompatible questions in a sequential way, such as `A and then B'." (Pothos and Busemeyer 2013, p. 257)
What we propose here is to explicitly pose these two questions.

\subsection{Four conjunction fallacy-like tasks}

In order to strengthen our experimental tests, we have considered four scenarios that have been shown in the literature to give rise to conjunction fallacies, from which we have built four experimental tasks --- a task consists for an agent in reading a text and then sequentially answering two yes-no questions.

The first task %, noted $T_{\text{Linda}}$, 
is drawn from the case of \textbf{Linda} (Tversky and Kahneman 1983):\footnote{Please refer to Appendix \ref{AppB} for the French version that was actually used in the experiments.}
\begin{itemize}
\item[--] Text: ``Linda is 31 years old, single, outspoken, and very bright. She majored in philosophy. As a student, she was deeply concerned with issues of discrimination and social justice, and also participated in anti-nuclear demonstrations."
\item[--] $Q_F$: ``According to you,%
\footnote{We have added the terms ``According to you" at the beginning of each question so that the agents do not begin wondering on the possible existence of a correct answer. Furthermore, these terms convey the spirit of the initial instructions of Tversky and Kahneman (1983) to evaluate a characteristic according to its probability, which supposes here a subjective part of judgment.} 
is Linda a feminist?"
\item[--] $Q_B$: ``According to you, is Linda a bank teller?"
\end{itemize}

The second task %, noted $T_{\text{Bill}}$, 
is drawn from the case of \textbf{Bill} (Tversky and Kahneman 1983): 
\begin{itemize}
\item[--] Text: ``Bill is 34 years old. He is intelligent, but unimaginative, compulsive, and generally lifeless. In school, he was strong in mathematics but weak in social studies and humanities."
\item[--] $Q_A$ : ``According to you, is Bill an accountant?"
\item[--] $Q_J$ : ``According to you, does Bill play jazz for a hobby?"
\end{itemize}

The third task %, noted $T_{\text{Mr. F.}}$, 
is drawn from the case of \textbf{Mr. F.} (Tversky and Kahneman 1983):
\begin{itemize}
\item[--] Text: ``A health survey was conducted in a representative sample of adult males in France of all ages and occupations. Mr. F. was included in the sample. He was selected by chance from the list of participants."
\item[--] $Q_H$:  ``According to you, has Mr. F. already had one or more heart attacks?"
\item[--] $Q_M$: ``According to you, is Mr. F. over 55 years old?"
\end{itemize}

The fourth task %, noted $T_{\text{K.}}$, 
is drawn from the case of \textbf{K.}, a Russian woman (Tentori et al. 2013):
\begin{itemize}
\item[--] Text: ``K. is a Russian woman".
\item[--] $Q_N$: ``According to you, does K. live in New-York?"
\item[--] $Q_I$: ``According to you, is K. an interpreter?"
\end{itemize}

So as to increase the robustness of our results, we have chosen these four tasks as they display different kinds of conjunction fallacies, in the sense of Tversky and Kahneman (1983) who have distinguished between M--A and A--B paradigms. In the former, a model M (the text describing the person) is positively associated with an event A (one of the two sentences forming the conjunction) and negatively with the other event B. This is the case of the Linda scenario: the introductory text M is positively associated with the event ``Linda is a feminist" and negatively with the other one ``Linda is a bank teller". Also, Bill's scenario is of type M--A. Differently, in the A--B paradigm, A is positively associated with B, but not with the model M. For instance, ``Mr. F. is over 55 years old" is positively associated with ``Mr. F. already had one or more heart attacks", but not with the text. The scenario of the Russian woman seems to correspond to neither paradigm: the positive association occurs between the text M and the conjunction of the two constituents A and B, and not with only one of them, so we might call it M--(AB) --- the fact that the woman is Russian is strongly associated with the fact that she lives in New York and is also an interpreter.

\subsection{Experimental protocol}

Conjunction fallacies and quantum-like models have been studied by scholars of various fields, and in particular by psychologists and economists (cf.~Section~\ref{sec_intro}). To keep with these two traditions,  we have chosen not to limit ourselves to one experimental protocol --- which also has the advantage of increasing the robustness of the experimental findings. We have varied the administration method, with paper questionnaires like in the psychological tradition and with computer implementations like in the economical tradition, with and without payment.

We have carried out three experiments (cf. Table~\ref{tab_listExp} for a summary). 
In the first experiment, two tasks were successively presented to the subjects: that of Mr. F. and that of Bill. The experiment was conducted in March and April 2015 at the University of Tours and of Nice Sophia Antipolis (France), with a total of 496 students in medicine, economics and management. In the psychological tradition, the tasks were implemented with paper questionnaires, in the lecture hall at the end of classes. Because of the improvised recruitment without appointment, and because of the short length of the task, the students were not paid, like in the psychological tradition. These tasks are noted $T_{\text{Mr. F.}}^{p}$ and $T_{\text{Bill}}^{p}$, with an index ``$p$" for ``paper".

The second experiment successively featured the 4 tasks introduced above in the following order: K. the Russian woman, Mr. F., Bill and Linda. The experiment was conducted on April 2015 at the LAMETA, the experimental economics Laboratory of the University of Montpellier~1 (France), in 19 sessions, with a total of 302 students possibly from any discipline. 
In the economics tradition, the tasks were implemented on computers (created with the z-Tree program, Fischbacher 2007), and students were recruited online and received a show-up fee (5 or 9 euros, according to their campus of origin) to remunerate their attendance and to reduce the effect of selection bias. 
These tasks are noted  $T_{\text{K.}}^{c, \, \text{\euro}}$, $T_{\text{Mr. F.}}^{c  , \, \text{\euro}}$, $T_{\text{Bill}}^{c, \, \text{\euro}}$ and $T_{\text{Linda}}^{c, \, \text{\euro}}$, with an index ``$c$" for ``computer" and a euro for the payment.

A third experiment involved the task of Linda, in a mixed methodology. It was conducted on October 2014 in the LEEN, the experimental economics laboratory of the University Nice Sophia Antipolis, with a computerized questionnaire. 354 students were recruited on the fly at the end of the classes, and were not paid for the short task. This task is noted $T_{\text{Linda}}^{c}$, with an index ``$c$". % and a tilde for the absence of payment.

\begin{table}[!tb]
\centering
\caption{Experimental tasks that were carried out, together with their administration methods, the location, and the number of subjects involved. Two dashed horizontal lines separate into three groups the seven experimental tasks, corresponding to three distinct experiments.\label{tab_listExp}}
\begin{tabular}{llllll}\\
\textbf{Exp.} &  \textbf{Task ID}  & \textbf{Scenario} & \textbf{Administration}&  \textbf{Location} & \textbf{No. of subjects} \\
\hline
1 & $T_{\text{Mr. F.}}^{p}$ 			&  Mr. F.  					& Paper, not paid     		&  Nice, Tours & 496 \\
 & $T_{\text{Bill}}^{p}$ 				&  Bill          				& Paper, not paid    		&  Nice, Tours & 496 \\
\hdashline[0.5pt/4pt]
2 & $T_{\text{K.}}^{c, \, \text{\euro}}$ 			&  K. & Computer, paid 			& Montpellier & 302 \\
& $T_{\text{Mr. F.}}^{c, \, \text{\euro}}$ 				&	Mr. F.         			& Computer, paid 			&  Montpellier & 302 \\
& $T_{\text{Bill}}^{c, \, \text{\euro}}$ 				& Bill 				   		& Computer, paid 			&  Montpellier & 302 \\
& $T_{\text{Linda}}^{c, \, \text{\euro}}$ 				& Linda  					& Computer, paid 			&  Montpellier & 302 \\
\hdashline[0.5pt/4pt]
3 & $T_{\text{Linda}}^{c}$	& Linda         			& Computer, not paid 	&  Nice        & 354 \\
\end{tabular}
\end{table}

Each task comes in two treatments, according to the ordering of the questions $Q_X$ and $Q_Y$.
According to a between-subject approach which is consistent with the literature on question order effect, each subject only receives one treatment of a task: either $Q_X$ then $Q_Y$, noted $(Q_X, Q_Y)$, or $Q_Y$ then $Q_X$, noted $(Q_Y, Q_X)$.
We took all necessary precautions to organize the sessions in such a way as to avoid discussions among students having and having not performed the experiment, and we ensured that the students had never heard of the Linda story nor studied order effect or conjunction fallacy.

All experimental sessions were run in compliance with the ethical rules of the LEEN and of the LAMETA. These rules are known by subjects when they enrol on the web-based recruitment platform. Even in the experimental sessions run at the end of classes in the lecture hall, confidentiality and anonymity of data collection were guaranteed. Students participated on a voluntarily basis and they were informed about the nature of the experimentation.

An objection to our protocol has to be considered. In our first two experiments, several tasks are successively presented to a same subject. Is not there a risk that a former task influences the answers provided to the following task(s)? Two considerations enable to answer negatively. Firstly, from an experimental perspective, Stolartz-Fantino et al. (2003) proposed six conjunction fallacy tasks in sequence and observed no significant difference in conjunction error rate over the tasks. So, there seems to be no learning effect or influence between tasks. Secondly, the quantum-like models themselves imply theoretically that the tasks do not have any influence on one another. This is so because the stories, and in particular the mental representations that the subjects form of them, are sufficiently distant from each other, in a technical quantum-mechanical sense:  the basis vectors of the different tasks (Linda is feminist, Bill plays jazz for a hobby, ...) are compatible in the quantum mathematical framework, which implies that no order effect can occur among the different tasks (see e.g. Wang and Busemeyer 2013).
It might be empirically the case that our tasks \textit{do} influence one another, but no matter: as here we only intend to test these quantum-like models, and not to establish experimental results that could be used outside of these models, we are justified in relying on them for our protocol. Quantum-like models justify our experimental protocol that tests them, and that is sufficient.

\section{Experimental outcomes} \label{expOutcomes}
% \usepackage{multirow}

This section presents the experimental outcomes for each task.
As a reminder, with $Q_X$ and $Q_Y$ denoting the two questions of a task, $(Q_X, Q_Y)$ denotes the treatment where $Q_X$ is posed first and $Q_Y$ is posed second, and $(Q_Y, Q_X)$ the treatment in the reverse order. Two response categorical variables $\mathbf{X}$ and $\mathbf{Y}$ are introduced. $\mathbf{X} \in \{ X_y, X_n \}$ is the Bernoulli random variable represented by question X assuming two possible values $X_y$ for ``yes" and $X_n$ for ``no". Similarly, $ \mathbf{Y} \in \{ Y_y, Y_n\} $ is the Bernoulli random variable represented by question Y assuming values $Y_y$ for ``yes" and $Y_n$ for ``no".
Both treatments $(Q_X,Q_Y)$ and $(Q_Y,Q_X)$ are thus statistical experiments described by multinomial distributions. For each task and treatment, there are four possible outcomes, for instance for the $(Q_X, Q_Y)$ treatment: $\{(X_y, Y_y), (X_n, Y_y), (X_y, Y_n), (X_n, Y_n)\}$. The joint [relative] frequency of people responding $i$ to the first question $Q_X$ and then $j$ to the second question $Q_Y$ is noted $n[f](X_i, Y_j)$.
Table~\ref{tab-contingence} reports the joint [relative] frequencies for each treatment, for our seven tasks. \setlength{\tabcolsep}{0.25em}
\begin{table}[!tb]
\centering
\caption{Cross tabulations of the joint [relative] frequencies $n[f](X_i, Y_j)$ for the two treatment of the seven tasks. The four possible outcomes of the $(Q_X, Q_Y)$ treatment are reported in the upper part of the table, while the four possible outcomes of the $(Q_Y, Q_X)$ treatment are reported in the bottom part.
}
\label{tab-contingence}
\vspace{0.2cm}
\begin{tabular}{l|ll;{0.5pt/4pt}ll;{0.5pt/4pt}ll;{0.5pt/4pt}ll;{0.5pt/4pt}ll;{0.5pt/4pt}ll;{0.5pt/4pt}ll} %{l|cccccccccccccc}
\textbf{Task} & \multicolumn{2}{c;{0.5pt/4pt}}{\textbf{$T_{\text{Mr. F.}}^{p}$}} & \multicolumn{2}{c;{0.5pt/4pt}}{\textbf{$T_{\text{Bill}}^{p}$}} & \multicolumn{2}{c;{0.5pt/4pt}}{\textbf{$T_{\text{K.}}^{c, \, \text{\euro}}$}} & \multicolumn{2}{c;{0.5pt/4pt}}{\textbf{$T_{\text{Mr. F.}}^{c, \, \text{\euro}}$}} & \multicolumn{2}{c;{0.5pt/4pt}}{\textbf{$T_{\text{Bill}}^{c, \, \text{\euro}}$}} & \multicolumn{2}{c;{0.5pt/4pt}}{\textbf{$T_{\text{Linda}}^{c, \, \text{\euro}}$}} & \multicolumn{2}{c}{\textbf{$T_{\text{Linda}}^{c}$}}\\
$(Q_X, Q_Y)$ & \multicolumn{2}{c;{0.5pt/4pt}}{$(Q_M, Q_H)$} & \multicolumn{2}{c;{0.5pt/4pt}}{$(Q_A, Q_J)$} & \multicolumn{2}{c;{0.5pt/4pt}}{$(Q_N, Q_I)$} & \multicolumn{2}{c;{0.5pt/4pt}}{$(Q_M, Q_H)$} & \multicolumn{2}{c;{0.5pt/4pt}}{$(Q_A, Q_J)$} & \multicolumn{2}{c;{0.5pt/4pt}}{$(Q_B, Q_F)$} & \multicolumn{2}{c}{$(Q_B, Q_F)$} \\ \hline
$n[f](X_y, Y_y$)            &	56	&	[0.21]	&	22  &	 [0.10]  & 14	&	[0.09]	&	17	&	[0.12]	&	9	&	[0.06]	&	1	&	[0.01]	&	7	&	[0.04]	\\
$n[f](X_y, Y_n$)            &	10	&	[0.04]	&	100 &	 [0.46]  & 12	&	[0.08]	&	7	&	[0.05]	&	83	&	[0.55]	&	1	&	[0.01]	&	6	&	[0.04]	\\
$n[f](X_n, Y_y$)            &	72	&	[0.28]	&	31  &	 [0.14]  & 18	&	[0.12]	&	34	&	[0.23]	&	9	&	[0.06]	&	104	&	[0.67]	&	86	&	[0.51]	\\
$n[f](X_n, Y_n$)            &	123	&	[0.47]	&	65  &	 [0.30]  & 111	&	[0.72]	&	89	&	[0.61]	&	49	&	[0.33]	&	49	&	[0.32]	&	68	&	[0.41]	\\                                          
\hdashline[0.5pt/4pt]                                                                                                                                   
\multicolumn{1}{r|}{total} &	261	&		&	218 &  & 155  &		&	147	&		&	150	&		&	155 &		&	167	&		\\                                                               
%\end{tabular}
%\begin{tabular}{l|ll;{0.5pt/4pt}ll;{0.5pt/4pt}ll;{0.5pt/4pt}ll;{0.5pt/4pt}ll;{0.5pt/4pt}ll;{0.5pt/4pt}ll}
$(Q_Y, Q_X)$ & \multicolumn{14}{c}{}\\
\hline
$n[f](Y_y, X_y$)            &	27	&	[0.11]	&	26  &	 [0.09]  & 10	&	[0.07]	&	15	&	[0.10]	&	20	&	[0.13]	&	4	&	[0.03]	&	5	&	[0.03]	\\
$n[f](Y_n, X_y$)            &	19	&	[0.08]	&	163 &	 [0.59]  & 8	&	[0.05]	&	15	&	[0.10]	&	80	&	[0.53]	&	2	&	[0.01]	&	8	&	[0.04]	\\
$n[f](Y_y, X_n$)            &	52	&	[0.22]	&	28  &	 [0.10]  & 24	&	[0.16]	&	34	&	[0.22]	&	12	&	[0.08]	&	79	&	[0.54]	&	57	&	[0.30]	\\
$n[f](Y_n, X_n$)            &	137	&	[0.58]	&	61  &	 [0.22]  & 105	&	[0.71]	&	91	&	[0.59]	&	40	&	[0.26]	&	62	&	[0.42]	&	117	&	[0.63]	\\                                               
\hdashline[0.5pt/4pt]                                                                                                                                   
\multicolumn{1}{r|}{total} &	235	&		&	278 &	  & 147  &		&	155	&		&	152	&		&	147	&		&	187	&		\\
\end{tabular}
\end{table}

\section{Statistical analysis and test of research hypotheses} \label{StatAnalysis}

To analyze the above experimental results, we proceed in two steps. The first step is technical: we perform the three statistical tests presented in Section~\ref{sec_empirical-tests} %on the GR equations, the order effect, and the QQ equality 
(Sections \ref{subsec_test-GR-equations} to \ref{subsec_test-QQ-equality}). In the second step, we take a more general viewpoint and we interpret the results of the tests in relation with several major research hypotheses (Section~\ref{subsec_interp-results}).

\subsection{Test of the GR equations}
\label{subsec_test-GR-equations}

The GR equation (\ref{eq_mix0011}, or equivalently \ref{eq_mix1001}, see Section~\ref{sec:GReq}) consists in the equality of 4 conditional probabilities. Thus, it is equivalent to 6 two-by-two equalities to be tested:  %We focus our statistical test on the GR equation~\ref{eq_mix0011}, being equivalent to the second set (GR eq.~\ref{eq_mix1001}). 
%The six tests (T1-T6) are then:
\begin{numcases}
\strut
\label{6Tests}
\text{T1:  } f(X_y|Y_y) = f(Y_y|X_y),\label{firstTest}\\
\text{T2:  } f(X_y|Y_y) = f(X_n|Y_n),\\
\text{T3:  } f(X_y|Y_y) = f(Y_n|X_n),\\
\text{T4:  } f(Y_y|X_y) = f(X_n|Y_n),\\
\text{T5:  } f(Y_y|X_y) = f(Y_n|X_n),\\
\text{T6:  } f(X_n|Y_n) = f(Y_n|X_n).\label{lastTest}
\end{numcases}
It is worth noting that the rejection of only one test is sufficient to state that a GR equation is not verified on a task.
We test all the equivalences with 6 statistical tests adopting conditional relative frequencies with the null hypothesis that the two conditional relative frequencies are equal (please refer to appendix \ref{AppA1} for a detailed description of the statistical test, taken from Boyer-Kassem et al. 2016).
Our two-tailed test imply that the null hypothesis of equality between the two conditional frequencies at the $K\%$ significance level is rejected if:
\begin{equation}
p\text{-value} = 2\cdot\left(1-{\rm CDF_{stdNorm}}\left(\left|\frac{\log ({\rm OR})}{\rm SE_{logOR}}\right|\right)\right)\leq \frac{K}{100}.
\end{equation}
${\rm CDF_{stdNorm}}$ is the cumulative distribution function of the standard normal distribution (mean = 0 and standard deviation = 1). $\rm log(OR)$ and $\rm SE_{logOR}$ are respectively the log odds ratio and its standard error. 
%As showed in appendix \ref{AppA1}, they are function exclusively of joint relative frequencies, which enables to use the information from Table \ref{tab-contingence} to perform the test. 
The multiple comparisons (the 6 simultaneous tests) and the joint testing of 7 tasks require performing a correction of the type I error, if we want to control for the probability of making at least one false discoveries in the whole table. We apply the Bonferroni correction, which is the most conservative one as it makes false positives much less liable to occur. We apply it doubly, on the 6 tests and on the 7 tasks. The risk is obviously to restrict our statistical inference to only one case by increasing the type II error, that is, the presence of false negatives, but the adoption of this correction guarantees that the conclusion of rejections that we provide is robust.
Accordingly, we adopt adjusted $p$-values as follows: 
\begin{equation}
\mbox{adjusted $p$-value} = 6 \cdot 7 \cdot p\text{-value}.
\end{equation}
Table \ref{statHyp1} reports adjusted $p$-values for each of the six tests. It shows that for all tasks, at least two out of the six statistical tests reject the null of equality between the two conditional relative frequencies. Hence, we can safely say that the GR equations are not empirically satisfied in our experiments.

\setlength{\tabcolsep}{1em}
\begin{table}[!tb]
\centering
\caption{Adjusted $p$-values for each task and test. The value is in bold when the null is rejected at the 5\% significance level. With the double Bonferroni corrections, the probability of having at least one false positive in the entire table is guaranteed to be less than 5\%.}
\begin{tabular}{ l | l l l l l c c}\label{statHyp1}
Task ID & \textbf{T1} & \textbf{T2}& \textbf{T3} & \textbf{T4} & \textbf{T5} & \textbf{T6} & \textbf{\#R at 5\%} \\
\hline
$T_{\text{Mr. F.}}^{p}$ & \textbf{0.00} & 0.07 & 21.58	& \textbf{0.00} &	\textbf{0.00} & \textbf{0.00} & 4\\
$T_{\text{Bill}}^{p}$ & \textbf{0.00}	&	\textbf{0.00}	& 2.57 & 0.84 & 0.14 & \textbf{0.00} & 3\\
\hdashline[0.5pt/4pt]
$T_{\text{K.}}^{c, \, \text{\euro}}$ & 2.61 & \textbf{0.01} & \textbf{0.00} & \textbf{0.00} & \textbf{0.00} & 4.11 & 4\\
$T_{\text{Mr. F.}}^{c, \, \text{\euro}}$ & 0.08 & 34.67 & 3.02 & \textbf{0.00} & \textbf{0.00} & 0.66 & 2\\
$T_{\text{Bill}}^{c, \, \text{\euro}}$ & \textbf{0.00} & \textbf{0.00} & \textbf{0.01} & 0.92 & 0.16 & \textbf{0.00} & 4\\
$T_{\text{Linda}}^{c, \, \text{\euro}}$ & 0.92 & 21.93 & 0.62 & \textbf{0.00} & \textbf{0.00} & \textbf{0.00} & 3\\
\hdashline[0.5pt/4pt]
$T_{\text{Linda}}^{c}$ & \textbf{0.01} & 21.05 & \textbf{0.00} & \textbf{0.00} & \textbf{0.00} & \textbf{0.00} & 5\\
\end{tabular}
\end{table}

\subsection{Test of the order effect}
\label{subsec_test-order-effect}

Consider now the test of the order effect. The tradition in the literature is to test the null of absence of order effect (e.g. Wang and Busemeyer 2013 and Wang et al. 2014). Table~\ref{tab_test-null-absence-OE} reports the adjusted $p$-values of the log-likelihood ratio test with a Bonferroni correction for such a test. The null is rejected in two tasks ($T_{\text{Mr. F.}}^{p}$ and $T_{\text{Linda}}^{c}$), which enables us to assert safely that these two tasks exhibit an order effect. It could be tempting to infer that five tasks out of seven do not exhibit an order effect. %, and hence that Hyp.~\#2 is rejected in all these cases. 
However, it is well-known that there are possible errors of type II, which in that case are not well controlled. As here we need to be able to say with a high confidence level whether there is \textit{no} order effect, this traditional test is insufficient. For that reason, we propose a more rigorous test, with the reverse null hypothesis that there exists an order effect. 
%This requires the adoption of a non-standard statistical test.

\setlength{\tabcolsep}{1em}
\begin{table}[!tb]
\centering
\caption{Adjusted $p$-values for each task. The value is in bold when the null of absence of order effect is rejected at the 5\% significance level. With the Bonferroni correction, the probability of having at least one false positive in the table is guaranteed to be less than 5\%.
%(est-ce que ça n'est pas faux de dire que la null hyp est rejetée \textit{sur telle tâche}, puisqu'on a affaire à une hyp nulle générale ?).
\label{tab_test-null-absence-OE}}
\vspace{0.2cm}\begin{tabular}{ l | c}
Task ID & absence of OE \\
\hline
$T_{\text{Mr. F.}}^{p}$ 			&          \textbf{0.01}	\\
$T_{\text{Bill}}^{p}$ 			&           0.25    \\
\hdashline[0.5pt/4pt]
$T_{\text{K.}}^{c, \, \text{\euro}}$ 			&         3.60	\\
$T_{\text{Mr. F.}}^{c, \, \text{\euro}}$ 	&          2.84 	\\
$T_{\text{Bill}}^{c, \, \text{\euro}}$ 			&          0.91	\\
$T_{\text{Linda}}^{c, \, \text{\euro}}$ 			&          0.50    \\
\hdashline[0.5pt/4pt]
$T_{\text{Linda}}^{c}$ 	&  \textbf{0.00}   \\
\end{tabular}
\end{table}

This reverse null hypothesis requires the adoption of a specific statistical test.
We choose the two one-sided test (TOST) procedure of equivalence testing for binomial random variables (Barker et al. 2001)\footnote{Equivalence tests are commonly adopted in medicine to state if novel therapies have equivalent efficacies to the ones currently in use. For instance they are used by FDA to establish the equivalence between a generic drug versus an established drug.}. Equivalence tests are used to assess whether there is a practical difference in two means of occurrence (binomial proportions). This concept is formalized by defining a constant $\delta$ called the equivalence margin, which defines a range of values for which the two means are ``close enough" to be considered equivalent. This arbitrary notion of ``close enough" is the most distinctive feature of equivalence testing.

Concretely, equivalence testing in our context amounts to considering as the null hypothesis $H_0$ that, in two distinct treatments $(Q_X, Q_Y)$ and $(Q_Y,Q_X)$, the absolute difference between two probabilities of occurrence of an event $e$, $p_{XY}(e)$ and $p_{YX}(e)$, is greater than a pre-specified level $\delta > 0$ (formally, $H_0(e) : |p_{XY}(e) - p_{YX}(e)|> \delta$). %The alternative hypothesis is $H_a(e) : |p_{XY}(e) - p_{YX}(e)|< \delta$ that the difference is less than $\delta$ and corresponds to the hypothesis that the probabilities are ``close enough" to be considered equivalent.
The order effect is commonly studied with respect to a specific answer to one of the questions, that is, $X_y$, $X_n$, $Y_y$ or $Y_n$. For instance, the order effect of the event ``answering yes to question $Q_X$" ($X_y$) is evaluated by estimating the absolute difference of the marginal probabilities (marginal relative frequencies) of the event $X_y$ in the two treatments $(Q_X,Q_Y)$ and $(Q_Y, Q_X)$, formally, $|p_{XY}(X_y) - p_{YX}(X_y)|$. According to our notations, $p_{XY}(X_y) = p(Y_y, X_y) + p(Y_n, X_y)$ and $p_{YX}(X_y) = p(X_y, Y_y) + p(X_y, Y_n)$. 
As $p(X_y)=1-p(X_n)$, the order effect of the event $X_y$ is equivalent to the order effect of the event $X_n$, for both treatments.
In order to state that there is no order effect, or that the order effect is insignificant in a task, it is necessary and sufficient to test the validity of the two null hypotheses $H_0(e_1)$ and $H_0(e_2)$ at a time for both questions $Q_X$ and $Q_Y$ simultaneously. The following set of equations should be verified:
\begin{eqnarray}
|p_{XY}(X_y) - p_{YX}(X_y)| = |p(Y_y, X_y) + p(Y_n, X_y) - p(X_y, Y_y) - p(X_y, Y_n)| &>& \delta, \\
|p_{XY}(Y_y) - p_{YX}(Y_y)| = |p(Y_y, X_y) + p(Y_n, X_y) - p(X_y, Y_y) - p(X_y, Y_n)| &>& \delta.
\end{eqnarray}
Statistically, we adopt the TOST procedure which is based on a confidence interval approach, that is, it declares the equivalence, at a chosen nominal value of significance $\alpha$, if a $(1 - 2\alpha)100\%$ equal-tailed confidence interval is completely contained in the interval $[-\delta, \delta]$. We consider the simple asymptotic interval approach to estimate the confidence interval
\begin{equation}
CI: f_{xy}(e) - f_{yx}(e) \pm Z_{\alpha}\cdot\sqrt{\frac{f_{xy}(e)(1-f_{xy}(e))}{n_{xy}(e)}+\frac{f_{yx}(e)(1-f_{yx}(e))}{n_{yx}(e)}},
\end{equation}
where $Z_{\alpha}$ represents the $(1-2\alpha)$100$^{th}$ percentile of a standard normal distribution and the notation $f(e)$ stands for the marginal relative frequency which is the estimator of the marginal probability $p(e)$. If the CI is contained in the interval $[-\delta, \delta]$, then we reject the null hypothesis.

\begin{figure}[!tb]
\begin{center}
\includegraphics[scale=0.35]{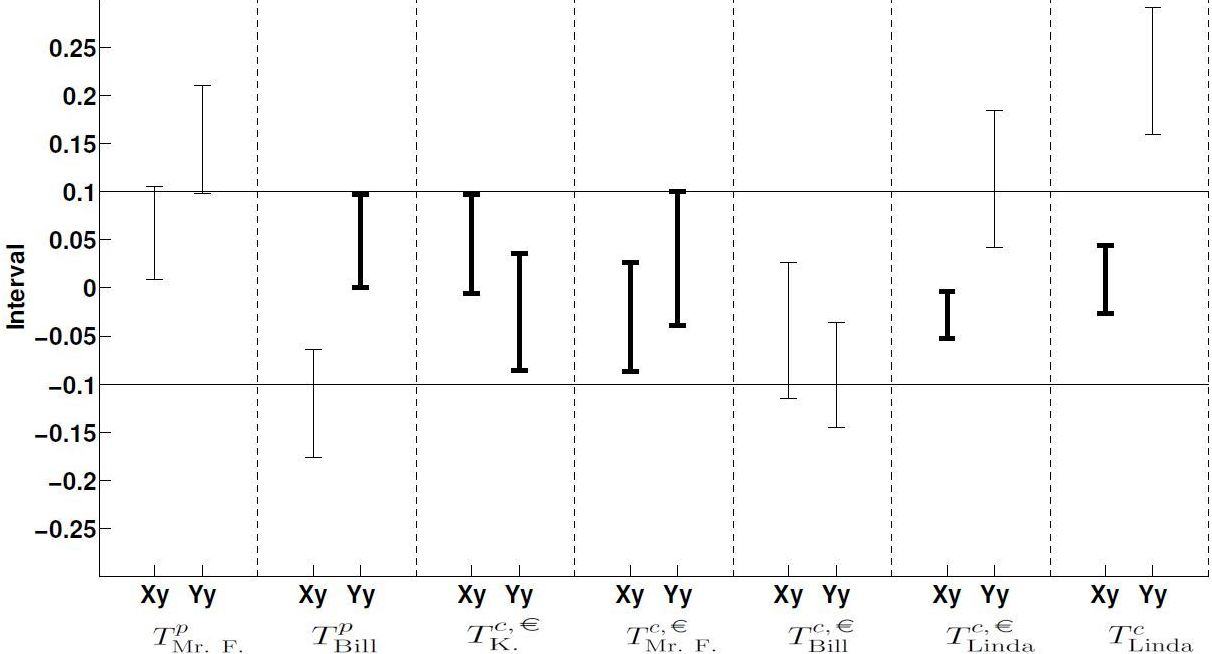}
\end{center}
\caption{Equivalence testing for the seven tasks and two events $X_y$ and $Y_y$. For each task, two vertical segments correspond to the estimated confidence interval (CI) for the events for the ``yes" answer to both questions $Q_X$ and $Q_Y$. Intervals in bold are entirely contained within the $\delta$ interval $\left[-0.1, 0.1\right]$ highlighted with two horizontal lines.\label{plotEquivalence}}
\end{figure}

Figure \ref{plotEquivalence} shows the results of the test for the seven tasks, with our choice of a nominal value of significance $\alpha=5\%$ and a threshold $\delta=0.1$.  
Before commenting on these results, let us justify the chosen values of the two parameters $\alpha$ and $\delta$.
A large value of $\delta$ easily leads to rejections, while a small value hardly leads to rejections (a value of $\delta = 0$ has no statistical meaning). In the TOST procedure, the $\delta$ value is supposed to be chosen before the experiment is run, from indications from the literature or from some \textit{a priori} consideration.\footnote{In medicine, the literature reports that \textit{ex ante} discussions among practitioners are required to find an agreement on the $\delta$ value to specify what the irrelevant differences are between the efficacies of similar drugs or medical treatments (see e.g. Walker and Nowacki 2011).}
In our case, there is no clear indication coming from the literature that bears on a similar problem (i.e. we could not find any work addressing the issue of testing the null of presence of order effect).
Yet, \textit{a priori} consideration can be attempted, as some theoretical studies provide simulated evidences of the power of the equivalence testing. Given similar statistical conditions, i.e. a sample size around 200 statistical units, $\delta = 0.1$ and $\alpha= 0.05$, the simulated power of the equivalence testing attains a probability value of around 0.75 of rejecting the null when the difference between the two relative frequencies is less than 0.05 (Barker et al. 2001, p.~282, Table~3). In other words, our choice of parameters enables to expect that, if we judge a difference of less or equal 0.05 to be irrelevant in terms of order effect, then the test is effective in three cases out of four.
Some \textit{a posteriori} justification of the value of $\delta$ can be added. 
Figure~\ref{plotEquivalence} shows a great variability in CIs between similar tasks, for instance between
$T_{\text{Mr. F.}}^{p}$ and $T_{\text{Mr. F.}}^{c, \, \text{\euro}}$, 
$T_{\text{Bill}}^{p}$ 	and $T_{\text{Bill}}^{c, \, \text{\euro}}$, or 
$T_{\text{Linda}}^{c, \, \text{\euro}}$ 	and $T_{\text{Linda}}^{c}$, 
and that variability (measured for instance as the difference of the top margin of both CIs) is of the order of 0.1. These pairs of tasks are not fully homogeneous in terms of administration method, but we think that it is sensible to consider them as highly informative of an inner variability of the order effect phenomenon, when the size of the sample is around 200 subjects.
Thus, it would not make much sense to choose a $\delta$ lower than that inner variability of 0.1. Our choice of 0.1 is thus the most conservative in this respect.

To strengthen the test, we also add the condition that the value 0 should be part of the CI.
Two out of the seven tasks ($T_{\text{K.}}^{c, \, \text{\euro}}$ and
$T_{\text{Mr. F.}}^{c, \, \text{\euro}}$) fulfill these two conditions: for both events $X_y$ and $Y_y$, the CIs are entirely contained within the $\delta$ interval $\left[-0.1, 0.1\right]$, and the value of $\delta=0$ is included in the estimated CI. Thus, these two tasks exhibit an order effect that can be deemed as insignificant.

Note that the results of our TOST test are in line with the more traditional test with the opposite null hypothesis reported above. In particular, the two tasks that do not exhibit an order effect according to the TOST test ($T_{\text{K.}}^{c, \, \text{\euro}}$ and $T_{\text{Mr. F.}}^{c, \, \text{\euro}}$) are exactly those which exhibit the highest adjusted $p$-values (Table~\ref{tab_test-null-absence-OE}), with a large margin compared to the other tasks.
This consistency is a clue that our choice of parameters $\alpha$ and $\delta$ are meaningful and not too permissive.

\subsection{Test of the QQ equality}
\label{subsec_test-QQ-equality}

To test the QQ equality, we adopt the statistical test proposed in Wang and Busemeyer (2013) and Wang et al. (2014), based on the log-likelihood ratio test, commonly used to compare the goodness of fit of two models.\footnote{We thank the authors of these two papers for kindly giving us their code to perform the statistical test.} %(please refer to Appendix \ref{AppA2} for a detailed description of the statistical test).\\
The two models are an unconstrained one and a constrained one by the QQ equality. The difference of the two log-likelihoods follows a $\chi^2$ statistic with degrees of freedom resulting from the difference of the degrees of freedom of each model.
As we perform the same test over 7 different tasks, we also adopt a Bonferroni correction of the type I error, which is the most conservative one. 
Table \ref{statHyp2} reports the adjusted $p$-values for each task, with the null hypothesis that the QQ equality is satisfied for all tasks.%
\footnote{It is worth noting that, when using the Bonferroni correction, the null hypothesis is here that \textit{all tasks} satisfy the QQ equality --- this is the so-called general null hypothesis. Even if we apparently reject only one task according to the adjusted $p$-values, we are in fact rejecting the corresponding general null hypothesis that all tasks are of the same nature. Somehow paradoxically, if we knew \textit{ex ante} that all tasks would be of different nature and that only some would satisfy the QQ equality, we should not adopt the Bonferroni correction but consider instead seven individual null hypotheses. The results of these individual tests without the Bonferroni correction would be that three out of the seven tasks are rejected (the $p$-values are easy to estimate from Table \ref{statHyp2} by dividing the adjusted $p$-values by 7). The rejected tasks correspond to three out of the four MA paradigm tasks, with Linda and Bill. This could be taken to suggest that there might be a difference between the nature of the tasks: MA tasks would tend to reject the QQ equality, whereas AB and (AB)M would tend to satisfy it. %, but because there is no order effect, as showed hereinafter. 
In any case, we prefer to keep the conservative Bonferroni correction to make unquestionable rejections, and we invite further research to study the individual hypothesis more in depth.}
It is clear that for only one task ($T_{\text{Linda}}^{c}$, last row) we can reject the null, thus stating that the QQ equality is not satisfied. 
Conversely, for all tasks except the last one, nothing can be concluded. They are either false negatives or cases where the QQ equality is satisfied.

\setlength{\tabcolsep}{1em}
\begin{table}[!tb]
\centering
\caption{Adjusted $p$-values for each task. The value is in bold when the null of satisfaction of the QQ equality is rejected at the 5\% significance level. With the Bonferroni correction, the probability of having at least one false positive in the table is guaranteed to be less than 5\%.
%(est-ce que ça n'est pas faux de dire que la null hyp est rejetée \textit{sur telle tâche}, puisqu'on a affaire à une hyp nulle générale ?).
\label{statHyp2}}
\vspace{0.2cm}\begin{tabular}{ l | c}
Task ID & QQ equality \\
\hline
$T_{\text{Mr. F.}}^{p}$ 			&          5.402	\\
$T_{\text{Bill}}^{p}$ 			&          0.324    \\
\hdashline[0.5pt/4pt]
$T_{\text{K.}}^{c, \, \text{\euro}}$ 			&          4.226	\\
$T_{\text{Mr. F.}}^{c, \, \text{\euro}}$ 	&          3.356	\\
$T_{\text{Bill}}^{c, \, \text{\euro}}$ 			&          6.200	\\
$T_{\text{Linda}}^{c, \, \text{\euro}}$ 			&          0.167    \\
\hdashline[0.5pt/4pt]
$T_{\text{Linda}}^{c}$ 	&  \textbf{0.001}   \\
\end{tabular}
\end{table}

\subsection{Interpretation of the results and relation with general research hypotheses}
\label{subsec_interp-results}

On the basis of the above experimental results, we now would like to test three research hypotheses that have motivated the quantum-like modeling literature on conjunction fallacy, and that correspond to the building blocks of the current models presented in Section~2. This shall provide some interpretation of the bare statistical results obtained in Sections \ref{subsec_test-GR-equations} to \ref{subsec_test-QQ-equality}.
The first two hypotheses have already been presented in the introduction and concern the validity of quantum-like models, while the third one is larger and goes beyond quantum-like models:
\begin{itemize}
\item \textbf{Hyp.~\#1:} Non-degenerate quantum-like models (presented in Section~2) can account for the conjunction fallacy,
\item \textbf{Hyp.~\#2:} Non-degenerate or degenerate quantum-like models (presented in Section~2) can account for the conjunction fallacy,
\item \textbf{Hyp.~\#3:} The conjunction fallacy account can rely on a question order effect account.
\end{itemize}

The first hypothesis is the simplest and less general one. It restricts accounts of the conjunction fallacy to the simplest versions of the quantum-like models, i.e. non-degenerate ones, where answers are represented by 1-D subspaces. This is the hypothesis made in Franco (2009), who only considers non-degenerate models. % (they are considered, among others, in all the other papers considered here, cf. Section~2). 
This hypothesis implies that the GR equations are empirically verified. 
As Section~\ref{subsec_test-GR-equations} has shown that the GR equations are never verified in our experiments, we can safely say that the first hypothesis is empirically refuted by our data. In other words, non-degenerate quantum-like models cannot account for order effects. This refutes the proposal by Franco (2009), who has only considered non-degenerate models --- all other quantum-like models cited in Section~2 are not refuted, since they also consider degenerate models.
The rejection of the first hypothesis echoes recent debates.  The empirical inadequacy of non-degenerate models for the conjunction fallacy has already been discussed, although the question had not been definitely settled (cf. Tentori and Crupi 2013 and Pothos and Busemeyer 2013, p.~315-316). In a similar vein, it has been shown that non-degenerate models for order effect are not empirically adequate (Boyer-Kassem et al. 2016). Overall, our result is in line with previous suggestions that degenerate models should be preferred to non-degenerate models, as the latter should be considered as ``toy models" only (e.g. Busemeyer and Bruza 2012, Busemeyer et al. 2015). %The rejection of Hyp~\#1 is in line with this stance.

The second research hypothesis extends the first one by considering also degenerate models, that is, models in which an answer is represented by a $N$-D subspace, e.g a plane. This hypothesis is shared by all papers cited in the beginning of Section~2, except Franco (2009): the conjunction fallacy can be accounted for by quantum-like models in general, be they non-degenerate or degenerate. As argued in Section~\ref{sec_empirical-tests}, non-degenerate and degenerate models have (i) to display an order effect and (ii) to respect the QQ equality. Thus, the second hypothesis is testable by means of the test of the order effect and that of the QQ equality. 
Table~\ref{statHyp2Final} summarizes the findings on these matters.
Both tests' results are reported, the satisfaction of the QQ equality in the second column and the presence of order effect in the third one. The last column reports the joint outcomes of the two tests, that is, the outcome of the logical operator ``and", because either one test or the other one are sufficient to refute the quantum-like models of conjunction fallacy considered in this paper.
Recall that we have adopted a very conservative approach on the error of type~I, so as to be conclusive with a high degree of certainty. So, we can be quite sure that the second research hypothesis is rejected in at least three out of seven tasks. 
Our conclusion here is that the quantum-like models cannot account for the general phenomenon of the conjunction fallacy. It is the first time that such a strong result is obtained experimentally.

\setlength{\tabcolsep}{1.2em}
\begin{table}[!tb]
\begin{center}
\caption{Statistical results for the second research hypothesis. A dash means an absence of pronouncement. \label{statHyp2Final}}
\vspace{0.2cm}\begin{tabular}{ l | c ;{0.5pt/4pt} c | c }
Task ID & \textbf{QQ} & \textbf{OE} & \textbf{QQ and OE}\\
\hline
$T_{\text{Mr. F.}}^{p}$ 			& - 	& Yes 	&  - 			\\
$T_{\text{Bill}}^{p}$ 			& - 	& -  	&  -  			\\
\hdashline[0.5pt/4pt]                    
$T_{\text{K.}}^{c, \, \text{\euro}}$ 				& -  	& No 	&  \textbf{No}  \\
$T_{\text{Mr. F.}}^{c, \, \text{\euro}}$			& -  	& No 	&  \textbf{No}  \\
$T_{\text{Bill}}^{c, \, \text{\euro}}$ 				& -  	& - 	&  -  			\\
$T_{\text{Linda}}^{c, \, \text{\euro}}$ 			& -  	& -  	&  -  			\\
\hdashline[0.5pt/4pt]                      
$T_{\text{Linda}}^{c}$ & No & Yes  	&  \textbf{No}  \\
\end{tabular}
\end{center}
\end{table}

The third hypothesis is not restricted to quantum-like models, but is concerned with the general idea that the conjunction fallacy is related to a question order effect between suitable questions (for instance in the Linda scenario between the questions $Q_L$ and $Q_F$). It implies that an order effect must be observed in our experiments, and thus this hypothesis is testable by means of the order effect test.
Two out of seven tasks exhibit no (or insignificant) order effect, as shown in Section~\ref{subsec_test-order-effect}. And yet, the corresponding scenarios (K. and Mr. F.) do exhibit a conjunction fallacy. % (cf. Section~\ref{sec_expdesign}). 
These results suggest that the third hypothesis, according to which the conjunction fallacy can be accounted for from an order effect, seems to be experimentally refuted. Note that the consequences of the rejection of this hypothesis have an even much broader impact than the ones deriving from the rejections of previous hypotheses: not only are we rejecting the original modeling strategy exploited by the quantum-like literature based on the introduction of an order effect to explain the conjunction fallacy, but we are also preventing its adoption for any other alternative theory (Bayesian, heuristics...). 
The conjunction fallacy cannot be reduced, in terms of mental acts, to the order effect phenomenon. This finding sheds some new light into an important modeling issue.

\section{Conclusion}
\label{sec_discussion}

We have considered the quantum-like accounts of the conjunction fallacy that have been proposed or defended by Franco (2009), Busemeyer et al. (2011), Busemeyer and Bruza (2012), Pothos and Busemeyer (2013) and Busemeyer et al. (2015) --- which common trait is to represent the belief of the decision-maker with the quantum state. We have tested three empirical predictions of these models: the GR equations (Boyer-Kassem et al. 2016), that apply to non-degenerate versions only of the models, the existence of an order effect and the QQ equality (Wang and Busemeyer 2013), which apply to both non-degenerate and degenerate versions of the models, hence to the most general version of the papers. Such tests cannot be performed in traditional conjunction fallacy experiments, in which subjects have to rank propositions, but require an order effect experiment, in which two yes-no questions are asked in either order. So, the tests concern empirical predictions that are not the data that the models were supposed to explain in the first place, but are predictions of the models anyway, and are directly related to the core feature of the models, namely the incompatibility between questions.
We have performed such order effect experiments, by using a robust protocol that varies the stories (Linda, Bill, Mr. F., K.), the administration method (paper questionnaires or computer), and a possible payment, with seven tasks in total and several hundreds of subjects. 

Our empirical results clearly reject the hypothesis that non-degenerate models can account for the conjunction fallacy (which is the hypothesis made in Franco 2009). This confirms the recent tendency from the advocates of the quantum-like approach to consider non-degenerate models as toy models only. 
Most importantly, our results also reject the more general hypothesis that non-degenerate or degenerate models can account for the conjunction fallacy, which is the hypothesis made in all other papers. 
As we have used very conservative statistical tests, we can safely say this general hypothesis is refuted in at least three tasks out of seven. So the present paper provides the first clear experimental rejection of the quantum-like explanation of the conjunction fallacy.

Now, it may be possible that not \textit{all} instances of the conjunction fallacy can be accounted for in a quantum-like fashion, but that \textit{some} instances can. For instance, our experimental results have not formally excluded that Bill's scenario could be amenable to a quantum-like account. There is room for possible future experimental research here --- a possible line of division to be investigated could be between AB and MA scenarios of conjunction fallacies. But thus, the quantum-like account would loose its generality, which was its strength. % (however, already not able to account for double conjunction fallacies)
Moreover, if quantum-like models were to apply to some cases of conjunction fallacies, it seems very likely that it should be degenerate versions, since non-degenerate one have been strongly ruled out. This comes with possible drawbacks or specific duties, as argued in Boyer-Kassem et al. (2016). In particular, a degenerate model resorts to some extra dimensions in the Hilbert space that should receive theoretical and experimental justifications so as not to be just \textit{ad hoc}. And more general tests on elementary dimensions can also be considered.

As our experimental results speak against the quantum-like models of the conjunction fallacy, they can be interpreted as indirect support in favor of alternative accounts of the conjunction fallacy, like Bayesian ones (e.g. Tentori et al. 2013), or other kinds of quantum-like models for the conjunction fallacy that have not been tested in this paper, like Yukalov and Sornette (2010, 2011). 
However, our results also provide some conclusions well beyond quantum-like modeling: they show that the conjunction fallacy cannot be accounted for by any model or mechanism that relies on order effect, or entails an order effect, between the two characteristics at play (``feminist" and ``bank teller" in Linda's case). Quantum-like models are well-known such examples, but it must be clear that any existing or future alternative explanation that involves a question order effect is ruled out. After the failure of quantum-like models, this places a hard constraint on alternative explanations of the conjunction fallacy. We suggest that future works should try to theoretically inquire whether alternative explanations predict an order effect, and to experimentally test it.

Even if the quantum-like models studied in this paper are not able to account for our data, a possible research strategy could be not to abandon the quantum-like modeling of the conjunction fallacy altogether, but instead to try to modify and improve it so that it finally agrees with the experimental data. In this spirit, one could investigate whether the use of a more general measurement theory or generalized observables could be adequate. For instance, the use of Positive Operator Valued Measures (POVMs), from quantum physics, has been recently applied to quantum-like models of cognition (cf. Khrennikov and Basieva 2014). However, it seems to face some new challenges like response replicability (cf. Khrennikov et al. 2014, Basieva and Khrennikov 2015). %, Khrennikov 2015). 
%Note that the new generalized quantum-like models should not only account for the existence of a conjunction fallacy, but also for the violation of the QQ equality and the absence of order effect in our experiments.
Another quantum-like line of research that does not face this problem considers a modification of the Born rule (Aerts and Sassoli de Bianchi, 2015).

As a last remark, our methodology has been here to test quantum-like models of the conjunction fallacy with new experimental predictions. We think this methodology could be fruitfully extended to quantum-like models that address other fallacies, such as the disjunction fallacy or the inverse fallacy.

\section*{Acknowledgments} 

Many thanks to Corrado Lagazio for suggestions on the statistical part. We would like to thank Jerome Busemeyer, Vincenzo Crupi, Dorian Jullien, Andrei Khrennikov, Katya Tentori, Vyacheslav Yukalov and participants at the 2015 conference ``Quantum Probability and the Mathematical Modelling of Decision Making" (Fields Institute, Toronto), for useful comments and suggestions. 
Experimental Economics laboratory at the University of Nice (LEEN - Nice Lab) and at the University of Montpellier (LAMETA) have kindly hosted the experiments. 
This work has benefited from the joint GIS RnMSHS and French CNRS grant ``Quantumtest".

\section*{References}

\newlength{\oldparindent}
\setlength{\oldparindent}{\parindent}

\setlength{\parindent}{-.8cm}
\begin{changemargin}{.8cm}{0cm}

\textsc{Aerts}, Diederik, Jan \textsc{Broekaert}, Marek \textsc{Czachor}  and Bart \textsc{D'Hooghe} (2011), ``A Quantum-Conceptual Explanation of Violations of Expected Utility in Economics", \textit{Quantum Interaction. Lecture Notes in Computer Science} 7052: 192-198.
%In: \textit{Proceedings of the Quantum Interaction}, Springer, p. 192--198.

%\textsc{Aerts}, Diederik, Liane \textsc{Gabora} and Sandro \textsc{Sozzo} (2013), ``Concepts and Their Dynamics: A Quantum-Theoretic Modeling of Human Thought", \textit{Topics in Cognitive Sciences} 5: 737--772.

\textsc{Aerts}, Diederik and Massimiliano \textsc{Sassoli de Bianchi} (2015), ``Beyond-Quantum Modeling of Question Order Effects and Response Replicability in Psychological Measurements", arXiv: 1508.03686.

\textsc{Aerts}, Diederik and Sandro \textsc{Sozzo} (2013), ``A Contextual Risk Model for the Ellsberg Paradox", \textit{Journal of Engineering Science and Technology Review} 4: 246--250.

\textsc{Aerts}, Diederik, Sandro \textsc{Sozzo}, and Jocelyn \textsc{Tapia} (2014), ``Identifying Quantum Structures in the Ellsberg Paradox", \textit{International Journal of Theoretical Physics} 53, 3666-3682.
%``A quantum model for the Ellsberg and Machina Paradoxes",  %\textit{Quantum Interaction. Lecture Notes in Computer Science}, 7620: 48--59.

\textsc{Allais}, Maurice (1953), ``Le comportement de l'homme rationnel devant le risque : critique des postulats et axiomes de l'école Américaine", \textit{Econometrica} 21: 503--546.

%\textsc{Ashby}, F. Gregory and Nancy A. \textsc{Perrin} (1988), ``Towards a unified theory of similarity and recognition", \textit{Psychological Review}, 95:124--50.

\textsc{Ashtiani}, Mehrad and Mohammad Abdollahi \textsc{Azgomi} (2015), ``A survey of quantum-like approaches to decision making and cognition", \textit{Mathematical Social Sciences} 75: 49--80.

\textsc{Atmanspacher}, Harald and Hartmann \textsc{Römer} (2012), ``Order effects in sequential measurements of non-commuting psychological observables", \textit{Journal of Mathematical Psychology} 56: 274--280.

\textsc{Barker}, Lawrence, Henry \textsc{Rolka}, Deborah \textsc{Rolka} and Cedric \textsc{Brown} (2001), ``Testing for Binomial Random Variables: Which Test to Use?", \textit{The American Statistician} 55 (4): 279--287.

\textsc{Basieva}  Irina and Andrei  \textsc{Khrennikov} (2014), ``On a possibility to combine the order effect with sequential reproducibility for quantum measurements", working paper, \href{http://arxiv.org/abs/1502.00132}{arXiv:1502.00132}.

\textsc{Birnbaum}, Michael H., Carolyn J. \textsc{Anderson} and Unda G. \textsc{Hynan} (1990), ``Theories of bias in probability judgment.", %in J.P. Caverni, J.M. Fabre, and M. Gonzalez (Eds), 
\textit{Cognitive biases: Advances in psychology} 68:477--499. % Amsterdam, the Netherlands: North Holland/ Elsevier.

\textsc{Bonini}, Nicolao, Katya \textsc{Tentori} and Daniel \textsc{Osherson} (2004), ``A different conjunction fallacy", \textit{Mind \& Language} 19 (2): 199--210.

\textsc{Boyer-Kassem}, Thomas, Sébastien \textsc{Duchêne}  and Eric \textsc{Guerci} (2016), ``Testing quantum-like models of judgment for question order effects",  \textit{Mathematical Social Sciences} 80 : 33-46.  
%GREDEG working paper 2015-06, \url{http://www.gredeg.cnrs.fr/working-papers.html}.

\textsc{Brandenburger}, Adam (2010), ``The relationship between quantum and classical correlation in games", \textit{Games and Economic Behavior} 69: 175--183.

\textsc{Bruza}, Peter D., Zheng J. Wang, and Jerome R. \textsc{Busemeyer} (2015), ``Quantum cognition: a new theoretical approach to psychology", \textit{Trends in Cognitive Sciences} 19(7): 383--393.

\textsc{Busemeyer}, Jerome R. and Peter D. \textsc{Bruza} (2012), \textit{Quantum models of Cognition and Decision}, Cambridge: Cambridge University Press.

\textsc{Busemeyer}, Jerome R., Marvin \textsc{Matthew} and Zheng \textsc{Wang} (2006a), ``A quantum information processing explanation of disjunction effects", in R. Sun and N. Miyake (ed.), \textit{Proceedings of the 28th Annual Conference of the Cognitive Science Society}, Erlbaum, % and the 5th International Conference of Cognitive Science}, 
p. 131–135.

\textsc{Busemeyer}, Jerome R., Emmanuel M. \textsc{Pothos}, Riccardo \textsc{Franco} and Jennifer S. \textsc{Trueblood} (2011), ``A Quantum Theoretical Explanation for Probability Judgment Errors", \textit{Psychological Review} 118 (2): 193--218.

\textsc{Busemeyer}, Jerome R., and Zheng \textsc{Wang} (2007), ``Quantum information processing explanation for interactions between inferences and decisions", in: D. Bruza, W. Lawless, K. van Rijsbergen and D.A. Sofge  (eds.), \textit{Quantum Interaction: Papers from the AAAI Spring Symposium, Technical Report SS-07-08}, AAAI Press, p. 91--97.

\textsc{Busemeyer}, Jerome R., Zheng J. \textsc{Wang}, and Ariane \textsc{Lambert-Mogiliansky} (2009), ``Empirical comparison of Markov and quantum models of decision making", \textit{Journal of Mathematical Psychology} 53: 423--433.

\textsc{Busemeyer}, Jerome R., Zheng J. \textsc{Wang}, Emmanuel M. \textsc{Pothos}, and Jennifer S. \textsc{Trueblood} (2015), ``The Conjunction Fallacy, Confirmation, and Quantum Theory: Comment on Tentori, Crupi, and Russo (2013)", \textit{Journal of Experimental Psychology: General} 144(1): 236--243.

\textsc{Busemeyer}, Jerome R., Zheng \textsc{Wang}, and James T. \textsc{Townsend} (2006b), ``Quantum dynamics of human decision-making", \textit{Journal of Mathematical Psychology} 50 (3): 220--241.

\textsc{Charness}, Gary, Edi \textsc{Karni}, and Dan \textsc{Levin} (2010), ``On the conjunction fallacy in probability judgment: New experimental evidence regarding Linda", \textit{Games and Economic Behavior} 68: 551--556.

\textsc{Conte}, Elio, Andrei Yuri \textsc{Khrennikov}, Orlando \textsc{Todarello}, Antonio \textsc{Federici}, Leonardo \textsc{Mendolicchio}, and Joseph P. \textsc{Zbilut} (2009), ``Mental States Follow Quantum Mechanics During Perception and Cognition of Ambiguous Figures", \textit{Open Systems \& Information Dynamics} 16(1): 1--17.

\textsc{Crupi}, Vincenzo, Branden \textsc{Fitelson}, and Katya \textsc{Tentori} (2008), ``Probability, confirmation and the conjunction fallacy", \textit{Thinking \& Reasoning} 14: 182--199.

\textsc{Danilov}, Vladimir I. and Ariane \textsc{Lambert-Mogiliansky} (2008), ``Measurable systems and behavioral sciences", \textit{Mathematical Social Sciences} 55(3): 315--340.

\textsc{Danilov}, Vladimir I. and Ariane \textsc{Lambert-Mogiliansky} (2010), ``Expected utility theory under non-classical uncertainty", \textit{Theory and Decision} 68(1): 25--47.

\textsc{Ellsberg}, Daniel (1961), ``Risk, ambiguity, and the Savage axioms", \textit{Quarterly Journal of Economics} 75(4): 643--669.

\textsc{Erceg}, Nikola and Zvonimir \textsc{Galic} (2014), ``Overconfidence bias and conjunction fallacy in predicting outcomes of football matches", \textit{Journal of Economic Psychology} 42 : 52--62.

\textsc{Fantino}, Edmund, James \textsc{Kulik}, Stephanie \textsc{Stolartz-Fantino}, and William \textsc{Wright} (1997), ``The conjunction fallacy: A test of averaging hypothesis", \textit{Psychonomic bulletin \& Review} 4(1): 96--101.

\textsc{Fischbacher}, Urs (2007), ``Z-tree : Zurich toolbox for ready-made economic experiments", \textit{Experimental Economics} 10(2): 171--178.

\textsc{Fisk}, John E. (2004), ``Conjunction fallacy", in R. F. Pohl (ed.), \textit{Cognitive illusions: A handbook on fallacies and biases in thinking, judgment, and memory}, London: Psychology Press.

\textsc{Franco}, Riccardo (2009), ``The conjunction fallacy and interference effects", \textit{Journal of Mathematical Psychology} 53: 415--422.

%\textsc{Gavanski}, Igor and David R. \textsc{Roskos-Ewoldsen} (1991), ``Representativeness and conjoint
%probability", \textit{Journal of Personality and Social Psychology} 61:181--94.

%\textsc{Gigerenzer}, Gerd (1991), ``How to make cognitive illusions disappear: Beyond `heuristics and biases' ", \textit{European Review of Social Psychology} 2: 83--115.

\textsc{Gigerenzer}, Gerd (1996), ``On narrow norms and vague heuristics: A reply to Kahneman and Tversky (1996)", \textit{Psychological Review} 103(3): 592--596.

\textsc{Haven}, Emmanuel and Andrei \textsc{Khrennikov} (2013), \textit{Quantum Social Science}, Cambridge: Cambridge University Press.

\textsc{Hartmann}, Stephan and Wouter \textsc{Meijs} (2012), ``Walter the banker: the conjunction fallacy reconsidered", \textit{Synthese} 184: 73--87.

\textsc{Hertwig}, Ralph (1997), ``Judgment under uncertainty: Beyond probabilities", Berlin, Germany: Max Planck Institute for Human Development. Unpublished manuscript.

\textsc{Hertwig}, Ralph, Björn \textsc{Benz} and Stefan \textsc{Krauss} (2008), ``The conjunction fallacy and the many meanings of `and' ", \textit{Cognition} 108: 740--753.

\textsc{Hertwig}, Ralph, and Valerie M. \textsc{Chase} (1998), ``Many Reasons or Just one: How Response Mode Affects Reasoning in the Conjunction Problem", \textit{Thinking \& Reasoning} 4(4): 319--352.

\textsc{Hertwig}, Ralph, and Gerd \textsc{Gigerenzer} (1999), ``The `Conjunction Fallacy' Revisited: How Intelligent Inferences Look Like Reasoning Errors" \textit{Journal of behavioral decision making} 12: 275--305.

\textsc{Hintikka}, Jaakko (2004), ``A Fallacious Fallacy", \textit{Synthese} 140: 25--35.

\textsc{Jarvstad}, Andreas, and Ulrike \textsc{Hahn} (2011), ``Source reliability and the conjunction fallacy", \textit{Cognitive Science} 35(4): 682--711.

\textsc{Kahneman}, Daniel (2011), \textit{Thinking, Fast and Slow,} New York: Farrar, Straus and Giroux. 

\textsc{Kahneman}, Daniel, and Amos \textsc{Tversky} (1996), ``On the reality of cognitive illusions: A reply to Gigerenzer's critique", \textit{Psychological Review} 103: 582--591.

\textsc{Khrennikov}, Andrei (2010), \textit{Ubiquitous Quantum Structure. From Psychology to Finance}, Heildelberg: Springer.

%\textsc{Khrennikov}, Andrei (2015), ``Quantum-like model of unconscious–conscious dynamics", \textit{Frontiers in Psychology} 6:997.

\textsc{Khrennikov}, Andrei and Irina \textsc{Basieva} (2014), ``Quantum model for psychological measurements: from the projection postulate to interference of mental observables represented as positive operator valued measures",  \textit{NeuroQuantology} 12:324-336.

\textsc{Khrennikov}, Andrei, Irina \textsc{Basieva}, Ehtibar N. \textsc{Dzhafarov}, and Jerome R. \textsc{Busemeyer} (2014), ``Quantum Models for Psychological Measurements: An Unsolved Problem", \textit{PLoS ONE} 9(10): e110909.

\textsc{Khrennikov}, Andrei and Emmanuel \textsc{Haven} (2009), ``Quantum mechanics and violations of the sure-thing principle: The use of probability interference and other concepts", \textit{Journal of Mathematical Psychology} 53: 378--388.

%\textsc{Krumhansl}, Carol L. (1978), ``Concerning the applicability of geometric models to similarity data: The interrelationship between similarity and spatial density", \textit{Psychological Review} 85: 445--63.

\textsc{Lambert-Mogiliansky}, Ariane,  Shmuel \textsc{Zamir} and Hervé \textsc{Zwirn} (2009), ``Type indeterminacy: A model of the KT(Kahneman–Tversky)-man", \textit{Journal of Mathematical Psychology} 53:349--361.

\textsc{Landsburg}, Steven E. (2004), ``Quantum game theory", \textit{Notices of the American Mathematical Society} 51(4): 394--399.

%\textsc{Landsburg}, Steven E. (2011), ``Quantum game theory", in: \textit{Wiley Encyclopedia of Operations Research and Management Science} 51(4): 394--399.

\textsc{Massaro}, Dominique W. (1994), ``A pattern recognition account of decision making", \textit{Memory \& Cognition } 22: 616--627.

\textsc{Mellers}, Barbara, Ralph \textsc{Hertwig} and Daniel \textsc{Kahneman} (2001), ``Do frequency representations eliminate conjunction effects? An Exercise in Adversarial Collaboration", \textit{Psychological Science} 12: 269--275.

%\textsc{Moore}, David W. (2002), ``Measuring new types of question-order effects", \textit{Public Opinion Quarterly}, 66: 80--91.

\textsc{Moro}, Rodrigo (2009), ``On the nature of the Conjunction Fallacy", \textit{Synthese} 171: 1--24.

\textsc{Nilsson}, Hâkan, and Patric \textsc{Andersson} (2010), ``Making the seemingly impossible appear possible: Effects of conjunction fallacies in evaluations of bets on football games", \textit{Journal of Economic Psychology} 31: 172--180.

\textsc{Nilsson}, Hâkan, Anders \textsc{Winman}, Peter \textsc{Juslin}  and Göran G. \textsc{Hansson} (2009), ``Linda is not a bearded lady : Configural weighting and adding as the cause of extension errors", \textit{Journal of Experimental Psychology: General} 138: 517--534.

\textsc{Oechssler}, Jörg, Andreas \textsc{Roider} and Patrick W. \textsc{Schmitz} (2009), ``Cognitive abilities and behavioral biases", \textit{Journal of Economic Behavior \& Organisation} 72: 147--152.
%\textsc{Peres}, Asher (1993), \textit{Quantum Theory: Concepts and Methods}, Kluwer.

%\textsc{Peijnenburg}, Jeanne (2012),  ``Case of confusing probability and confirmation", \textit{Synthese}, 184, 101--107.

\textsc{Piotrowski}, Edward W.  and Jan \textsc{Stadkowski} (2003), ``An Invitation to Quantum Game Theory", \textit{International Journal of Theoretical Physics} 42(5): 1089--1099.

\textsc{Pothos}, Emmanuel M.  and Jerome R. \textsc{Busemeyer} (2009), ``A quantum probability explanation for violations of ``rational" decision theory", \textit{Proceedings of the Royal Society B} 276:2171--2178.

\textsc{Pothos}, Emmanuel M.  and Jerome R. \textsc{Busemeyer} (2011), ``A quantum probability explanation for violations of symmetry in similarity judgments.", \textit{Proceedings of the 32nd Annual conference of the Cognitive Science Society} 2848--2854.

\textsc{Pothos}, Emmanuel M.  and Jerome R. \textsc{Busemeyer} (2013), ``Can quantum probability provide a new direction for cognitive modeling?", \textit{Behavioral and Brain Sciences} 36: 255--327.

%\textsc{Schuman}, Howard and Stanley \textsc{Presser} (1981), \textit{Questions and answers in attitude surveys: Experiments on question form, wording, and context}, Academic Press.

\textsc{Schupbach}, Jonah N. (2012), ``Is the conjunction fallacy tied to probabilistic confirmation", \textit{Synthese} 184: 13--27.

\textsc{Shafir}, Eldar B., Edward E. \textsc{Smith} and Daniel N. \textsc{Osherson} (1990), ``Typicality and reasoning fallacies", \textit{Memory \& Cognition} 18:229--239.

%\textsc{Shafir}, Eldar B., and Amos \textsc{Tversky} (1992), ``Thinking through uncertainty: nonconsequential reasoning and choice", \textit{Cognitive Psychology} 24:449--74.

\textsc{Shogenji}, Tomoji (2012), ``The degree of epistemic justification and the conjunction fallacy", \textit{Synthese} 184: 29--48.

\textsc{Stolarz-Fantino}, Stepanie, Edmund \textsc{Fantino}, Daniel J. \textsc{Zizzo}, and Julie \textsc{Wen} (2003), ``The conjunction effect: New evidence for robustness", \textit{American Journal of Psychology} 116(1):15--34.

\textsc{Tentori}, Katya, Nicolao \textsc{Bonini}, and Daniel \textsc{Osherson} (2004), ``The conjunction fallacy: a misunderstanding about conjunction", \textit{Cognitive Science} 28: 467--477.

\textsc{Tentori}, Katya and Vincenzo \textsc{Crupi} (2008), ``On the conjunction fallacy and the meaning of \textit{and}, yet again: A reply to Hertwig, Benz, and Krauss (2008)", \textit{Cognition} 122: 123--134.

\textsc{Tentori}, Katya and Vincenzo \textsc{Crupi} (2012), ``How the conjunction fallacy is tied to probabilistic confirmation: Some remarks on Schupbach (2009)", \textit{Synthese} 184: 3--12.

\textsc{Tentori}, Katya and Vincenzo \textsc{Crupi} (2013), ``Why quantum probability does not explain the conjunction fallacy", commentary in Pothos and Busemeyer (2013), p. 308-310.

\textsc{Tentori}, Katya, Vincenzo \textsc{Crupi} and Selena \textsc{Russo} (2013), ``On the Determinants of the Conjunction Fallacy: Probability Versus Inductive Confirmation", \textit{Journal of Experimental Psychology: General} 142(1): 235--255.

%\textsc{Tourangeau}, Roger, Lance J. \textsc{Rips} and Kenneth A. \textsc{Rasinski} (2000),  \textit{The psychology of survey response}, Cambridge University Press.

%\textsc{Tversky}, Amos (1977), ``Features of similarity", \textit{Psychological Review} 84(4): 327--52.

\textsc{Tversky}, Amos and Daniel \textsc{Kahneman} (1982), ``Judgments of and by representativeness", in D. Kahneman, P. Slovic
\& A. Tversky (eds), \textit{Judgment under uncertainty: Heuristics and biases}, Cambridge University Press, p. 84--98.

--- (1983), ``Extensional versus intuitive reasoning: The conjunction fallacy in probability judgment", \textit{Psychological Review} 90(4): 293--315.

\textsc{von Sydow}, Momme (2011), ``The Bayesian logic of frequency-based conjunction fallacies", \textit{Journal of Mathematical Psychology} 55: 119--139.

\textsc{Walker}, Esteban and Amy S. \textsc{Nowacki} (2011), ``Understanding Equivalence and Noninferiority Testing", \textit{Journal of General Internal Medicine} 26(2): 192-196.

\textsc{Wang}, Zheng J. and Jerome R. \textsc{Busemeyer} (2013), ``A quantum question order model supported by empirical tests of an a priori and precise prediction", \textit{Topics in Cognitive Science} 5(4): 689--710.

\textsc{Wang}, Zheng J., Tyler \textsc{Solloway}, Richard M. \textsc{Shiffrin} and Jerome R. \textsc{Busemeyer} (2014),  ``Context effects produced by question orders reveal quantum nature of human judgments", \textit{Proceedings of the National Academy of Science} 111(26): 9431-9436. 
%\url{www.pnas.org/cgi/doi/10.1073/pnas.1407756111}.

%\textsc{Wedell}, Douglas H., and Rodrigo \textsc{Moro} (2008), ``Testing boundary conditions for the conjunction fallacy: Effects of response mode, conceptual focus, and problem type", \textit{Cognition} 107: 105--136.

\textsc{Yukalov}, Vyacheslav I. and Didier \textsc{Sornette} (2010), ``Mathematical Structure of Quantum Decision Theory", \textit{Advances in Complex Systems} 13: 659--698.

\textsc{Yukalov}, Vyacheslav I. and Didier \textsc{Sornette} (2011), ``Decision theory with prospect interference and entanglement", \textit{Theory and Decision} 70: 283--328.

\end{changemargin}

\setlength{\parindent}{\oldparindent}

%\newpage

\appendix

\section{Testing the equality of two conditional relative frequencies}\label{AppA1}
The statistical test is to compare two conditional relative frequencies $y$ and $x$, with the null hypothesis that they are equal. The test is therefore
\begin{equation}
y=x \label{yxTest},
\end{equation}
where both $y$ and $x$ are observed as conditional relative frequencies.\\
Testing equation \ref{yxTest} is equivalent to test
\begin{equation*}
\log \left(\frac{y}{1-y} \right)=\log \left(\frac{x}{1-x} \right),
\end{equation*}
given that $y$ and $x$ are not equal to zero.\\
Alternatively, we can formulate the test in terms of the log odds ratio (OR)
\begin{equation*}
\log({\rm OR}) = \log \left(\frac{\frac{y}{1-y}}{\frac{x}{1-x}} \right) = 0.
\end{equation*}
Consider the first statistical test,
\begin{equation}\label{condProbTest}
\mbox{T1}: f(X_y|Y_y) = f(Y_y|X_y).
\end{equation}
We can thus test the following condition:
\begin{equation*}
\log\left(\frac{f(X_y| Y_y)}{1- f(X_y|Y_y)}\right) = \log\left(\frac{f(Y_y| X_y)}{1-f(Y_y|X_y)}\right), 
\end{equation*}
or
\begin{equation*}
\log\left(\frac{f(X_y| Y_y)}{f(X_n|Y_y)}\right) = \log\left(\frac{f(Y_y| X_y)}{f(Y_n|X_y)}\right).
\end{equation*}
By expressing the conditional relative frequencies in terms of joint frequencies, that is,
\begin{equation*}  
f(X_y|Y_y) = \frac{n(Y_y, X_y)}{n(Y_y, \cdot)}, \ f(Y_y|X_y) = \frac{n(X_y, Y_y)}{n(X_y, \cdot)}, \dots
\end{equation*}
with ${n(Y_y, \cdot)}$ and ${n(X_y, \cdot)}$ the $y$-components of the marginal frequencies of $\mathbf{Y}$ and $\mathbf{X}$, we obtain
\begin{equation*}
\log\left(\frac{n(Y_y, X_y)}{n(Y_y, \cdot)}\frac{n(Y_y, \cdot)}{n(Y_y, X_n)}\right) = \log\left(\frac{n(X_y, Y_y)}{n(X_y, \cdot )}\frac{n(X_y, \cdot)}{n(X_y, Y_n)}\right),
\end{equation*}
or simplifying
\begin{equation} \label{logORTest}
\log ({\rm OR}) = \log\left(\frac{n(Y_y, X_y)n(X_y, Y_n)}{n(Y_y, X_n)n(X_y, Y_y)}\right) = 0,
\end{equation}
We can thus test indifferently eq.~\ref{condProbTest} or \ref{logORTest}.\\
Given condition \ref{logORTest}, to perform the statistical test we suppose here that 
\begin{equation}
\frac{\log ({\rm OR})}{\rm SE_{logOR}} \sim \mathcal{N}(0, 1),
\end{equation}
where ${\rm SE_{logOR}}$ is the standard error of the log odds ratio. It is estimated as the square root of the sum of the inverse of all the joint frequencies that are considered in the estimation of the ${\rm OR}$:\\
\begin{equation}
{{\rm SE_{OR}} = \sqrt{\dfrac{1}{n(Y_y, X_y)} + \dfrac{1}{n(X_y, Y_n)} + \dfrac{1}{n(Y_y, X_n)} + \dfrac{1}{n(X_y, Y_y)}}}.
\end{equation}\\

Finally, we also apply the continuity correction to the estimation of OR, because the normal approximation to the binomial is used, which is effective in particular for small values of $n(X_i,Y_j)$ or $n(Y_j, X_i)$:
\begin{equation}
\log\left(\frac{(n(Y_y, X_y)+0.5)(n(X_y, Y_n)+0.5)}{(n(Y_y, X_n)+0.5)(n(X_y, Y_y)+0.5)}\right) = 0.
\end{equation}\\

%\newpage

\section{French version of the tasks}\label{AppB}

\textbf{``Linda"}:
\begin{itemize}
\item[--] Text: ``Linda a 31 ans, elle est célibataire, franche, et très brillante. Elle est diplômée en philosophie. Lorsqu'elle était étudiante, elle se sentait très concernée par les questions de discrimination et de justice sociale et avait aussi participé à des manifestations anti-nucléaires."
\item[--] $Q_F$ : ``Selon vous, Linda est-elle féministe ?"
\item[--] $Q_B$ : ``Selon vous, Linda est-elle employée de banque ?"
\end{itemize}

\textbf{``Bill"}:
\begin{itemize}
\item[--] Text: ``Bill a 34 ans. Il est intelligent, mais n'a pas d'imagination, il est compulsif, et généralement plutôt éteint. À l'école, il était fort en mathématiques, mais faible dans les sciences humaines et sociales."
\item[--] $Q_A$ : ``Selon vous, Bill est-il comptable ?"
\item[--] $Q_J$ : ``Selon vous, Bill joue-t-il du jazz pour ses loisirs ?"
\end{itemize}

\textbf{``Mr. F."}:
\begin{itemize}
\item[--] Text: ``Une enquête de santé a été menée en France sur un échantillon représentatif d'hommes adultes de tous âges et de toutes professions. Dans cet échantillon, on a choisi au hasard Monsieur F.
\item[--] $Q_H$ :  ``Selon vous, Monsieur F. a-t-il déjà eu une ou plusieurs attaques cardiaques ?"
\item[--] $Q_M$ : ``Selon vous, Monsieur F. a-t-il plus de 55 ans ?"
\end{itemize}

\textbf{``K."}: 
\begin{itemize}
\item[--] Text: ``K. est une femme russe."
\item[--] $Q_N$ : ``Selon vous, K. vit-elle à New York ?"
\item[--] $Q_I$ : ``Selon vous, K. est-elle une interprète ?"
\end{itemize}

\end{document}